# QUANTUM ANTENNAS

G. Ya. Slepyan[1], S. Vlasenko[2] and D. Mogilevtsev[2]

[1]School of Electrical Engineering, Tel Aviv University, Tel Aviv 69978, Israel

[2]Institute of Physics, National Academy of Sciences of Belarus, Nezavisimosti Ave. 68, Minsk 220072, Belarus

## Abstract

Due to the recent groundbreaking developments of nanotechnologies, it became possible to create intrinsically quantum systems able to serve as high-directional antennas in THz, infrared and optical ranges. Actually, the quantum antennas, as devices shaping light on the level of single quanta, have already become the key elements in nanooptics and nanoelectronics. The quantum antennas are actively researched for possible implementations in quantum communications, quantum imaging and sensing, and energy harvesting. However, the design and optimization of these emitting/receiving devices are still rather undeveloped in comparison with the well-known methods for conventional radio-frequency antennas. This review provides a discussion of the recent achievements in the concept of the quantum antenna as an open quantum system emitting via interaction with a photonic reservoir. We focus on bridging the gap between quantum antennas and their macroscopic classical analogues. We also discuss the ways of quantum-antenna implementation for different configurations based on such materials, as plasmonic metals, carbon nanotubes, and semiconductor quantum dots.

# 1. INTRODUCTION: ANTENNA CONCEPTS

The development of nanotechnologies and the progress of quantum electromagnetics at the nanoscale[1-3] led to the transfer of the typical low-frequency concepts (such as lumped circuits and guiding waves) to the IR and optical frequency ranges. Quantumness has taken the firm place in modern nanoeletronics and nanophotonics, and has become crucial for nanoscale emitting/receiving devices. Antennas have become quantum.

In this review, we attempt to outline the concept of the quantum antenna as a device exploiting such features as discreteness of the emitters energy levels, non-classicality of quantum states and quantum correlations. We discuss the manifestations of the quantumness in antennas, the ways of exploiting it to one's advantage and their realization in practice.

## 1.1 Macroscopic and nanoantennas

The antenna is one of the most crucial components for wireless communication systems, radars, remote sensing, deep-space applications, both on airborne, radioastronomy and earth-based platforms. An antenna is a device, designed to radiate (transmitting antenna) or receive (receiving antenna) electromagnetic waves. In other words, the antenna is defined as a system component, which transforms the near electromagnetic field (which appears in the lumped electric circuits, transmission lines, waveguides, fibers, etc) to the far field, and vice versa. In a modern wireless system, the antenna acts as a directional element to accentuate the transmitted or received energy in some directions while suppressing it in the others. A good design of the antenna can relax system requirements and improve the overall system performance. The recent example is the smartphone, for which the overall broadcast reception can be improved using a high performance antenna. It is important to note, that there exists the legalized terminology with respect to antennas, which can be found, for example, in Reference[4]. Widespread interest in antennas is manifested in the outstanding collection of the published books describing the different aspects of the subject: fundamental investigations, measurements, description of special types of antennas with their applications. They are addressed to the different categories of specialists: theorists, experimentalists, and engineers [5–18].

The history of antennas dates back to Heinrich Rudolph Hertz, who demonstrated the first wireless electromagnetic system in 1886. He produced a spark at the wavelength of 4 m in the gap of a transmitting dipole which was then detected as a spark in the gap of a nearby loop. In 1901, Guglielmo Marconi sent radio signals over large distances. He performed the first transatlantic transmission from Poldhu in Cornwall, England, to St. John's, Newfoundland. His transmitting antenna consisted of 50 vertical wires in the form of a fan, connected to the ground through a spark transmitter. The wires were supported horizontally by a guy-wire between two wooden poles of 60 meters height. The receiving antenna at St. John's was a 200-meter wire pulled and supported by a kite. It must be noticed that Marconi used the term "antenna" in his Nobel Prize Lecture[19].

From Marconi's inception till the 1940's, the antenna technology was primarily centered on wire-related radiating elements (long wires, dipoles, helices, rhombuses, fans, etc.), which were used either as single elements or in arrays. The frequencies of their operation were up to about UHF. It was not until World War II that the modern antenna technology was launched, and new elements of the aperture type (waveguide apertures, horns, reflectors, etc.) were primarily introduced. An important contributing factor at this period of time was the invention and implementation of new types of microwave sources (klystron, magnetron, traveling-wave tube, etc.) with the frequencies of 1 GHz and above. At that time, antennas were totally passive devices and were limited in performance by their physical size: larger antennas and more elementary emitters were better.

In the early 1970's, the great advances in microelectronic technologies led to creation of the microstrip antennas [18]. This type of antennas is simple in fabrication, lightweight, inexpensive, low profile, and conformal to the surface. Microstrip antennas and arrays can be flush-mounted to metallic or dielectric surfaces. The operational disadvantages of microstrip antennas include low efficiency, narrow bandwidth, and strong limitations on the driving power. Major advances in millimeter and sub-millimeter wave antennas were made in the 1990's, including integrated antennas, where the active and passive non-linear circuits were combined with the radiating elements in one compact device. An important contributing factor at this period was the invention and implementation of new types of millimeter wave sources (semiconductor diodes and transistors). One of the simplest microantenna geometries, experimentally realized with the standard lithographic processes, consists of two rods separated by a small gap [20]. This geometry supports large and controllable enhancement of the field at the gap, opening the opportunities of exciting via semiconductor generators.

Due to the development of nanotechnologies, it became possible to create quantum emitting systems in which the electrons are spatially confined. To see a typical scale of the confinement, required to have the electrons interacting with the visible light, let us consider, for example, the wavelength of the visible light in vacuum in the green spectral range. This wavelength is about 500 nm (which corresponds to the energy of about 2.5 eV). So, the green photons would be able to interact with the spatially confined electrons and induce the transitions between the lowest and the first excited energy states, if the length scale of the electron confinement, i.e. the length of the box, were on the order of 1 nm [21]. The spatial electron confinement is typically encountered in such quantum systems, as quantum wires, quantum dots and the like, which are named 'quantum emitters'.

Radiofrequency antenna engineering benefits from the fact that metals at radio frequencies behave as nearly lossless. It makes legitimate the charge dynamics description in terms of a "free-electron gas". However, the losses start mounting as we move toward shorter wavelengths, and for the microwave regime (the THz domain) the losses already become a constraint to deal with. Also, in contrast to RF antennas that always appear as sufficiently large passive elements connected to a feeding circuit, optical antennas often appear as isolated structures with feeding via an external illuminating source (optical pumping).

Depending on the size of the antenna, quantum effects can be significant in the description of the antenna electrodynamics. The electronic conduction band can be taken as continuous at macroscopic scales, but it breaks up into discrete states when the dimensions are rather small. As a result, the conventional description through permittivity and permeability becomes no longer valid [21]. Many experiments have optically probed this quantum size effect [21]. It manifests itself as a shift and broadening of the resonances, in addition to the appearance of a fine structure, corresponding to the transitions between the discrete energy levels.

The antenna feeding was conventionally modeled as a phenomenological source of the external voltage. However, for antennas, exhibiting quantum effects, such feeding model cannot be readily applied. Moreover, dichotomy of passive and active elements is not suitable for the case anymore. Generally, the quantum antenna is, in fact, a set of quantum emitters interacting with themselves and with the field. The quantum emitters, arranged in the antenna, can interact by a number of specific mechanisms: dipole-dipole, tunneling, spin-spin, etc. As a result, one has quantum correlations between the antenna emitters even if one assumes the initial state of the antenna and the field as uncorrelated and classical (and some correlated states can be even non-radiating)[22–24]. So, the emitters also interact with the external source in a reciprocal way, and can potentially back-influence it.

Much of the work, laid the foundations for the quantum description of light-matter interaction, was carried out in the 1990's providing physically substantiated models for quantizing electromagnetic field in the different types of nanostructures[25-29]. The antenna dynamics was treated in terms of collective oscillations of the system as a whole. The quantized form of these collective matter oscillations, such as plasmons, polaritons, excitons, were found to be quasiparticles, with both having wave-like and particle-like behaviour, as expected for quantum excitations. The hybrid nature of the antenna excitations as `quasiparticles' makes them intriguing from the fundamental point of view, with many of their quantum properties still being unknown to a large extent. In addition, their potential is promising for strong coupling of light to emitter systems, such as Rabi-oscillations in a single atom and Rabi-waves in quantum dot chains, via highly confined fields at the gap of the antenna [2,3,30–32]. It offers new tools for the quantum control of light, which enables the implementation of such devices, as efficient single-photon sources and transistors.

### 1.2. Classical antenna basics

Qualitative understanding of the classical radiation mechanism can be obtained by considering a pulse source attached to an open-ended conducting wire connected to the ground through a load at its open end. When the wire is initially energized, the charges (free electrons) in the wire are set in motion by the forces created over the electric field. When the charges are accelerated in the source end of the wire and decelerated during the reflection from its ends, the radiated fields are produced at each end and along the remaining part of the wire[5]. The internal forces receive energy from the charge buildup as its velocity is reduced to zero at the ends of the wire. Therefore, the charge acceleration due to exciting an electric field and the deceleration due to impedance discontinuities or smooth curves of the wire are the mechanisms responsible for electromagnetic radiation. Both the current density and the charge density appear as source terms in Maxwell's equations, and, thus, this interpretation of radiation can be used to explain steady-state antenna emission[5].

Let us now consider the formation of radiation from a two-wire segment of a transmission line. We begin with the geometry of a lossless two-wire transmission line, as shown in **Figure 1a**. The motion of the charges creates traveling wave current of the same magnitude along each of the wires. When the current arrives at the end of each of the wires, it undergoes complete reflection (with keeping the magnitude and inverting the phase). As a result, a pure standing wave of a sinusoidal form (**Figure 1a**) is formed in each wire.

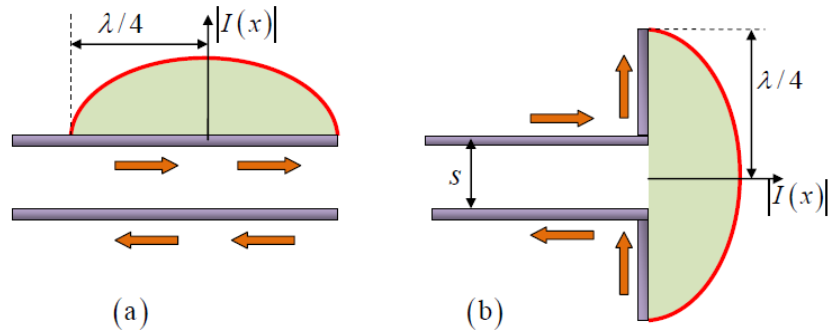

**Figure 1.** a) Open-ended transmission line excited by monochromatic current. b) Transformation of the transmission line to the dipole antenna excited by the transmission line via the gap at the center.

For the two-wire symmetrical transmission line, the current in a half-period of one wire is of the same magnitude but of the opposite phase compared to that in the corresponding half-period of the other wire. If the spacing between the two wires is very small compared with the wavelength ($s \ll \lambda$), the fields, radiated by the current of each wire, are essentially canceled by those of the other. The net result is an almost ideal nonradiating transmission line. However, the flared section of the transmission line can take the form shown in **Figure 1b**. This can be considered as a model of the widely used dipole antenna. Because of the standing-wave current distribution, it is also classified as a standing-wave antenna. If $l \leq \lambda/4$, the phase of the current standing wave in each arm is the same throughout its length. Next, it is oriented spatially in the same direction as that of the other arm (**Figure 1b**). Thus, the partial fields, radiated by the two arms, are identical to the radiation from a two-wire transmission line loaded by a linear dipole.

However, if we are moving to higher and higher frequencies in order to eventually end up with IR and visible light, the quantum effects in the antenna behavior will manifest itself more and more. The main difference between the interaction of low-frequency and very-high-frequency EM waves with condensed matter stems from two different mechanisms, which define the motion of charge carriers. The first one is tunneling over a potential barrier (for example, the electrons will tunnel with partial reflection at the turns of the wires at the angle 90 in **Figure 1b**). The second mechanism is discreteness of the energy levels. In this case, the condensed matter no longer behaves as a perfect conductor or a perfect insulator.

The space surrounding an antenna is usually subdivided into three regions: the reactive near-field region, the radiating near-field (Fresnel) region, and the far-field (Fraunhofer) region. These regions are designated in such a way to identify the field structure in each of them. The boundaries separating these regions are not unique, whereby the various criteria have been established and used for their identification. We quote definitions from Reference [4]. The reactive near-field region is defined as *"that region of the field immediately surrounding the antenna wherein the reactive field predominates."* For most antennas, the outer boundary of this region is commonly taken to exist at the distance $R < 0.62\sqrt{D^3/\lambda}$ from the antenna, where $D$ is the largest dimension of the antenna. The radiating near-field (Fresnel) region is defined as *"that*

*region of the field of an antenna between the reactive near-field region and the far-field region wherein radiation fields predominate and wherein the angular field distribution is dependent upon the distance from the antenna."* The radial distance at which this region exists, is $0.62\sqrt{D^3/\lambda} < R < 2D^2/\lambda$ (provided $D$ is large compared to the wavelength). The far-field (Fraunhofer) region is defined as "*that region of the field of an antenna where the angular field distribution is essentially independent on the distance from the antenna.*" In this region, the real part of the power density is dominant. The radial distance $R$ at which this region exists is $R > 2D^2/\lambda$ (provided $D$ is large compared to the wavelength). In this region, the field components are essentially transverse to the radial direction, and the angular distribution is independent on the radial distance.

The theory of macroscopic antennas is based on the classical electrodynamics and widely uses some fundamental properties of electromagnetic fields. The first one is the duality theorem, which follows from the invariance of Maxwell equations with respect to the exchanges $\mathbf{E} \to \mathbf{H}$, $\mathbf{H} \to -\mathbf{E}$, $\mathbf{J}_s^e \to -\mathbf{J}_s^m$, where $\mathbf{J}_s^m$ is the magnetic current density. Although magnetic sources are not physical, they are often introduced as electrical equivalents to simplify the solutions of physical boundary-value problems. For example, an aperture antenna, such as a waveguide or a horn, can be represented by either the equivalent magnetic current density or by the equivalent electric current density or both [33,34]. Another example is a small electric loop with the constant electric current, which is equal to the infinitesimal magnetic dipole [33].

Another fundamental principle is the reciprocity theorem [33,34]. Let us introduce a pair of electric and magnetic currents $\mathbf{J}_{1,2}^e, \mathbf{J}_{1,2}^m$, which produce the corresponding electromagnetic fields $\mathbf{E}_{1,2}, \mathbf{H}_{1,2}$. The fields and the currents are coupled via the relation

$$\int_V \left( \mathbf{E}_1 \cdot \mathbf{J}_2^e - \mathbf{H}_1 \cdot \mathbf{J}_2^m \right) dv = \int_V \left( \mathbf{E}_2 \cdot \mathbf{J}_1^e - \mathbf{H}_2 \cdot \mathbf{J}_1^m \right) dv \qquad (1)$$

where the volume $V$ is a sphere of an infinitesimally large radius. In the antenna theory, one of the applications of the reciprocity theorem is used to establish the fundamental relation between the properties of transmitting and receiving antennas. Let us take two identical antennas, 1 and 2, separated from each other. In one case, the antenna 1 is transmitting and is characterized by the set of sources $\mathbf{J}_1^e, \mathbf{J}_1^m$, produced by the generator. The antenna 2 is receiving, whereby it is characterized by the currents $\mathbf{J}_2^e, \mathbf{J}_2^m$, induced at the detector, placed inside the antenna. Now let us replace the generator with the detector, and vice versa. So, the transmitting antenna exchanges places with the receiving one. Treating these two antennas as a two-port network, with each port representing one of the antennas, it can be shown that the mutual (transfer) impedances between these two ports (antennas) are identical ($Z_{12} = Z_{21}$). This equality shows that the transmitting and receiving patterns of any antenna are identical, provided the antenna operates in a medium which is linear and isotropic. Notice that it is not always the case even in classical electrodynamics. A global perspective on electromagnetic nonreciprocity and the clarifications of the confusions that arose in the recent developments of the field are given in the recent review [35].

To analyze different types of antennas, one of the most efficient techniques is modeling of the source representing the actual antenna by the Field Equivalence Principle, often named Huygens-Fresnel principle[36]. With this method, the actual antenna is replaced by equivalent sources on the closed surface, surrounding the antenna producing the same fields as those radiated by the actual antenna. To insure the equivalence, the electric and magnetic current densities over the closed surface must be equal to

$$\mathbf{J}_s^e = \mathbf{n} \times \mathbf{H}_s \tag{2}$$

$$\mathbf{J}_s^m = -\mathbf{n} \times \mathbf{E}_s \tag{3}$$

where $\mathbf{n}$ is the external unit vector normal to the surface and $\mathbf{E}_s, \mathbf{H}_s$ are the electric and magnetic fields produced by the actual antenna on the chosen closed surface. Therefore, to form the currents by Equations (2) and (3), the tangential components of the magnetic and electric fields due to the actual antenna must be known over the chosen imaginary closed surface. The critical step here is to choose the imaginary closed surface so that the currents can be easily and intuitively approximated everywhere on it. The closed surface is often chosen so that the current densities are finite over a limited part of the surface and vanish elsewhere.

The electric and magnetic vector potentials at the far-field region are read as

$$\mathbf{A}^e = \frac{\mu_0}{4\pi r} e^{ikr} \int_S \mathbf{j}_s^e e^{-ikr'\cos\Psi} ds' = \frac{\mu_0}{4\pi r} e^{ikr} \mathbf{F}^e \tag{4}$$

$$\mathbf{A}^m = \frac{\varepsilon_0}{4\pi r} e^{ikr} \int_S \mathbf{j}_s^m e^{-ikr'\cos\Psi} ds' = \frac{\varepsilon_0}{4\pi r} e^{ikr} \mathbf{F}^m \tag{5}$$

where $\varepsilon_0, \mu_0$ are the vacuum permittivity and permeability, respectively [5–17,33,34]. The field radiation patterns

$$\mathbf{F}^{e,m} = e^{ikr} \int_S \mathbf{j}_s^{e,m} e^{-ikr'\cos\Psi} ds' \tag{6}$$

depend only on the angular variables.

A pattern of emitted field amplitudes is usually comprised of a number of lobes[5]. An extremely narrow radiation pattern, as well as its electrical control requirements, is usuallyunreachable by a single antenna element. To design the antennas with very large directivity, it is usually necessary to increase the antenna dimensions. This can be accomplished by increasing the electrical dimensions of the chosen single element. An alternative way to achieve large directivities without increasing the size of the individual elements, is to use an antenna array[5]. With arrays, it is practical not only to synthesize almost any desired amplitude radiation pattern, but the main lobe can be scanned by controlling the relative phase excitation between the elements. They find wide applications, for example a triangular array of dipoles used as a sectorial base-station antenna for mobile communication[5].

There are some parameters, which characterize the performance of an antenna system. The definitions of the parameters can be found in Reference [4], and the most important of them will be given below.

An antenna pattern is defined as a graphical representation of one of the antennas, usually in the far-field region. For the complete description, the parameters of interest are usually plotted as a function of the spherical angles $\theta, \varphi$. Parameters of interest include the amplitude, the intensity (energy), the phase, the polarization, and the directivity. The amplitude and energy patterns are usually comprised of a number of lobes.

(i) The main (major) lobe is defined as *"a radiation lobe containing the direction of maximum radiation. In certain antennas, such as multi-lobed or split-beam antennas, there may exist more than one major lobe"*.

(ii) The side lobe is defined as *"a radiation lobe in any direction other than that of the major lobe."* The amplitude level of the side lobe relative to the main lobe (usually expressed in decibels) is referred to as the side lobe level.

(iii) The input impedance is defined as *"the impedance presented by an antenna at its terminals"*. It is expressed at the terminals as the ratio of the voltage to the current or the ratio of the appropriate components of the electric to magnetic fields, and it is usually complex.

(iv) The radiation efficiency is defined as *"the ratio of the total power radiated by an antenna to the net power accepted by an antenna from the connected transmitter"*.

(v) The radiation intensity U is defined as *"the power radiated from an antenna per unit solid angle (steradian)"*. The radiation intensity is determined in the far field, and it is related to the real part of the power density by

$$U = \frac{r^2}{2} \operatorname{Re}\left(\mathbf{E} \times \mathbf{H}^*\right) \tag{7}$$

where $r$ is the spherical radial distance.

(vi) The isotropic radiator is defined as *"a hypothetical, lossless antenna having equal radiation intensity in all directions"*. Although such an antenna is idealization, it is often used as a convenient reference to define the directive properties of actual antennas. Its radiation intensity, defined as $U_0 = P_r / 4\pi$, where $P_r$ is the power radiated by the antenna.

(vii) The directivity is defined as *"the ratio of the radiation intensity in a given direction from the antenna to the radiation intensity averaged over all the directions"*. It can be written as

$$D = \frac{U(\theta,\varphi)}{U_0} = \frac{4\pi U(\theta,\varphi)}{P_r} \tag{8}$$

where $U(\theta,\varphi)$ is the radiation intensity in the direction $\theta, \varphi$. The directivity is an indicator of the relative directional properties of the antenna. As defined by Equation (8), the directional properties of the antenna are compared to those of an isotropic radiator.

(viii) The gain is defined as *"the ratio of the radiation intensity in a given direction, to the radiation intensity that would be obtained if the power accepted by the antenna were radiated isotropically"*.

$$G = \frac{U(\theta,\varphi)}{U_a} = \frac{4\pi U(\theta,\varphi)}{P_a} \tag{9}$$

where $P_a$ is the accepted (input) power to the antenna. If the direction is not specified, the direction of the maximum radiation (maximum gain) is implied.

## 2. Quantum antennas

We consider an antenna as "quantum" when it is necessary to apply the quantum formalism to describe it, i.e. to consider the coherent exchange of excitation between the energy levels or the modes of the considered system. So, to specify the cases when it is necessary, we define the emitting antenna structure in the following way: the feeding system, the spatially localized emitter system and the system of modes of the radiating reservoir. The feeding system excites the emitter system, the emitters emit into the modes of the radiation reservoir, and the modes carry the excitation away from the emitters.

The necessity to apply quantum description may arise when the feed creates a non-classical state in the emitter system. The system of emitting harmonic oscillators can be described classically if being excited in coherent states, or states representable as a positive-weight superposition of coherent projectors. However, the excitation of this very system, for example, to be in a Fock state or a squeezed state requires the full quantum description of the system dynamics. For the emitters with a finite number of energy levels, the quantum description is generally required, albeit there are approximations allowing the reduction of the level of complexity. Non-classicality of emitters state generally leads to the necessity of providing the quantum description of the set of modes representing the radiation reservoir. However, in this case, one can reduce the level of complexity, as well, if the quantities of interest are specified, for example, some correlation functions of the field.

It should be noted that it is not always easy to clearly distinguish between the feeding system, the emitters and the reservoir. Structuring of the reservoir may lead to strong coupling between the emitters and some reservoir modes (as is the case, for example, for the emitters in resonant cavities or waveguides), and so one would rather include these modes into the emitters structure. Feeding the emitters with other emitters may lead to the excitation exchange between them and to the necessity to consider both the feed and the antenna as a single system.

In order to proceed with more clearness for specifically quantum features of the considered antennas, and, at the same time, not to lose sight of a connection to classical antennas, here we limit ourselves mostly with the arrangements of quantum emitters, resembling systems of classical radiating dipoles, and we do not consider compound modal-emitters quantum structures.

### 2.1. The simplest quantum antenna: a two-level atom

To clarify the distinction between the parts of the quantum antenna, let us consider the simplest intrinsically quantum set-up: just a single two-level atom emitting into the vacuum of the reservoir of the electromagnetic field modes. This constitutes a fundamental model of the spontaneous emission, considered in standard textbooks on quantum optics (look, for example, at the classical text of Scully & Zubairy [2]).

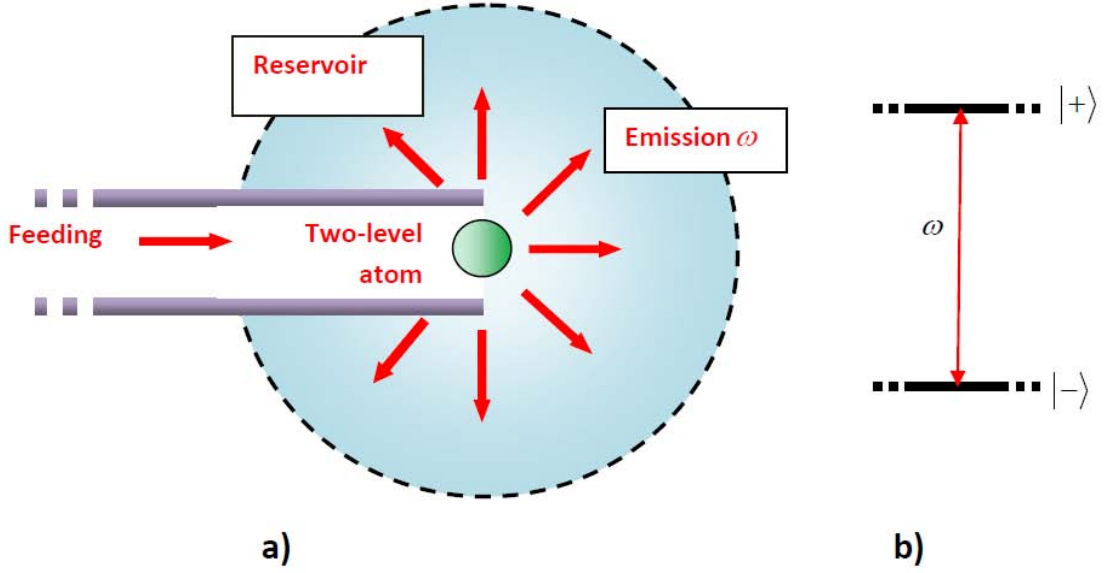

**Figure 2**. A schematic illustration of the simplest quantum antenna. a) Two-level single-atomic emitter placed at the open end of the feeding transmission line. The emitted energy goes to the photonic reservoir due to the "antenna-reservoir" coupling; b) Two-level emitter: the ground and the excited states separated by the frequency $\omega$.

For the homogeneous space as the reservoir of the field modes, this model describes the natural emission line width, gives the exponential decay law for the upper-state population dynamics and its rate, and connects this rate with the density of the reservoir modes by the famous "Fermi's golden rule"[37].

Here, we rewrite the traditional model in such a way as to emphasize the problems and the approaches arising for the general quantum-antenna scheme. We describe the whole feed-emitter-reservoir system as a sum of the following Hamiltonian parts,

$$\hat{H} = \hat{H}_{feed} + \hat{V}_{feed-emitter} + \hat{H}_{emitter} + \hat{V}_{emitter-reservoir} + \hat{H}_{reservoir}, \qquad (10)$$

where the Hamiltonians $\hat{H}_{name}$ describe free dynamics of the object "*name*", and the Hamiltonians $\hat{V}_{name1-name2}$ describe the interactions between the objects "*name1*" and "*name2*". Generally, all three parts of the antenna are quantum systems. They should be considered as a single compound object described by the wave function evolving according to the Schrodinger equation. All three parts may be entangled in the process of mutual interaction, and some action on the part of the system would affect the other parts.

Let us clarify the conditions under which all three parts of the system can be considered separately. First of all, if the coupling between the emitter and the field is weak, one can recourse to the dipole and the "rotating wave" approximations[2] and represent $\hat{V}_{emitter-field}$ as the process of a single excitation exchange between the emitter and the field,

$$\hat{V}_{emitter-field} = \hbar \sum_{\forall k}(v_k \hat{\sigma}^+ \hat{a}_k + h.c.), \qquad (11)$$

where the index $k$ conventionally numbers the reservoir field modes (distinct by the wavevectors and polarization) described by the bosonic annihilation and creation operators $\hat{a}_k$ and $\hat{a}_k^+$ and the operator $\hat{\sigma}^{\pm} = |\pm><\mp|$, the vectors $|\pm>$ describing the upper and the lower levels of the two-level system (TLS). The interaction constants $v_k$ are defined by the product of the emitter dipole moment vector and the vector amplitude of the $k$-th electric field eigenmode of the reservoir. These eigenmodes are found for the case as the solutions of the Maxwell equations for the elementary volume of space. The interaction constants $v_k$ describe how the emitter "sees" the reservoir.

The free parts for the emitters and the reservoir are $\hat{H}_{emitter} + \hat{H}_{field} = \hbar\omega\hat{\sigma}^+\hat{\sigma}^- + \sum_{\forall k} w_k \hat{a}_k^+ \hat{a}_k$

,with $\omega$ being the TLS transition frequency, and $w_k$ is the frequency of the $k$-mode.

In the absence of the feed part, with just a single excitation, the dynamics of the system is given by the wave function

$$|\psi(t)>= c_0(t)|+> |vacuum> + \sum_{\forall k} c_k(t)|-> |1>_k, \qquad (12)$$

where the vector $|vacuum>$ describes the vacuum state of the field, the $|1>_k$ is the state of the field with just one photon in the $k$-th mode, and the functions $c_x(t)$ are the scalar amplitudes. The state (12) shows the entanglement between the TLS and the field, and, generally, a field measurement will affect the state of the emitter.

However, in a vast majority of practical situations, the reservoir behaves like an inducer of loss, just taking the excitation away from the TLS and more or less homogeneously distributing this excitation between the modes. The excitation is not returned to the TLS, and the rate of the TLS state change (i.e. $dc_0(t)/dt$ in Equation (12)) is defined only by the TLS state (i.e. $c_0(t)$ in Equation (12)) at the moment. This constitutes the Markovian approximation[2,38]. Formally, the condition of the Markovian approximation applicability is the possibility to approximate the reservoir correlation function $K(\tau) = \sum_{\forall k}|v_k|^2 exp\{i(\omega - w_k)\tau\}$ with the delta-function of $\tau$.

The Fourier transform of the reservoir correlation function defines the projected local density of the state (PLDOS), $\varsigma(r, d, \omega)$, which is proportional to the decay rate of the TLS upper level population,

$$|c_0(t)|^2 \approx exp\{-\gamma t\}, \quad \gamma \propto \varsigma(r, d, \omega). \qquad (13)$$

So, while the Markovian approximation holds, one can safely neglect the effects of quantum correlations with the reservoir and consider the quantum dynamics only for the emitter, describing the reservoir influence by a few scalar parameters, such as the decay rate and the Lamb shift for the considered TLS case[38].

One should note two important features of the PLDOS. The first is the classical character: the PLDOS is defined from the solution of the classical Maxwell equations. The second is the possibility of describing the emission dynamics with the help of the Markovian approximation and the PLDOS for a huge variety of the reservoirs, including the one structured at the nanoscale level. Actually, even for the emitters inside the photonic crystal structures [64] (and generally in nanophotonics [40-42], including optical antennas[43] and, specifically, single-photon nanoantennas[44]), one is still pretty much within the region of the Markovian approximation validity. Here, the reservoir influence can be described by the change in the spontaneous emission rate plus, additionally, by the frequency shift experienced by the TLS. It must be noticed that even in these cases, the interaction between the emitter and the reservoir modes can be quite strong, and the induced frequency shifts can be large enough to significantly affect the bandwidth of the emitter radiation[45]. One can even seek to manipulate the emission spectrum and, at the same time, greatly enhance the emission rate, as is the case for the recently suggested hybrid antenna-cavity scheme[44,46].

The Markovian approximation generally fails when the reservoir structure separates a few field modes, strongly interacting with the TLS. An archetypal case of such a reservoir structure is the TLS inside a high-quality single-mode resonator described by the famous "Jaynes-Cummings model"[47]. There, the field remains strongly correlated with the TLS through all the dynamics, and the reservoir correlation function is just a harmonic function. A more realistic situation with the antenna is, for example, a quantum emitter near the surface of a small particle supporting a discrete set of resonant surface modes. The emitter can be strongly coupled to these modes (for example, see the comparatively recent work[48] on a quantum emitter in the vicinity of a nanowire). General prescription for considering these kinds of situations is to include all the discrete modes, strongly interacting with the emitter, into a compound antenna structure, and treat it as a single object emitting into the environment. Curiously, by increasing the spectral width (the leakage rate) of the quantized mode interacting with the TLS, one eventually arrives into the region of the Markovian approximation validity[49–51]. The strongly leaking modes again "submergeg" into the reservoir and become a part of it.

It should be noticed that the concept of the "non-responding", or "Markovian" reservoir is not really simple. Recently, it has become a subject of intensive studies, aimed to investigate a flow of correlations and information between the object and the reservoir, and to provide a strict definition of "Markovianity"[52,53]. However, in our consideration here, we universally assume the simplest empirical notion of "Markovianity" as a possibility to reduce the reservoir correlation function to the delta-function.

Under such "Markovianity", the expression for the positive-frequency part of the emitted field in the far-field zone from the TLS emitter, has the following simple form[2]:

$$\hat{\mathbf{E}}^{(+)}(\mathbf{r}, t) \propto \frac{[\mathbf{r} \times [\mathbf{d} \times \mathbf{r}]]}{|\mathbf{r}|^3} \hat{\sigma}^-(t) exp\left\{-i\frac{\omega}{c}\frac{\mathbf{r}}{|\mathbf{r}|}\mathbf{R}\right\}, \qquad (14)$$

where **r** is the observation point, and **R** is the position of the TLS. The different time arguments in the left and right parts of Equation (14) take account of the delay incurred for the field propagating from the point **R** to **r**.

Equation (14) immediately points to the manifestation of the quantumness in our simplest single-emitter antenna: for the initially completely excited TLS, $|\psi(0)>= |+> |vacuum>$, the average amplitude of the emitted field is always zero. Even when there is initial atomic coherence, $\langle \hat{\sigma}^-(0) \rangle \neq 0$, one generally has $\langle \hat{\mathbf{E}}^{(-)}(\mathbf{r},t)\hat{\mathbf{E}}^{(+)}(\mathbf{r},t)\rangle \neq |\langle \hat{\mathbf{E}}^{(+)}(\mathbf{r},t)\rangle|^2$. However, the distribution of the emitted field intensity will be still the same as for the classical emitting dipole.

The concept of the two-level quantum emitter leads to the fundamental model of the quantum dipole antenna, which generalizes the conventional classical model for the quantum case. It corresponds to a quantum emitter placed inside or in the vicinity of a nanowire supporting a suitable type of a quasiparticle (surface plasmon, exciton-polariton, etc.). Spontaneous emission of the excited atom creates a quasiparticle guided toward the edges of the wire. The scattering by the edges leads to the emission of photons. Such type of antennas has been experimentally implemented [54–56] (some details of the experiments are considered in Section 5).

The operator of the emitted field in the far-field zone of the antenna is

$$\hat{\mathbf{E}}^{(+)}(\mathbf{r},t) = \mathbf{e}_\theta \frac{\omega_0^2 d_{ab}}{4\pi\varepsilon_0 c^2} \int_{-l/2}^{l/2} \frac{1}{|\mathbf{r}-\mathbf{r}'(x')|} f(x') \hat{\sigma}^-\left(t - \frac{|\mathbf{r}-\mathbf{r}'(x')|}{c}\right) dx' \quad (15)$$

where $\mathbf{e}_\theta$ is the unit vector of the spherical system with the origin at the emitter, $d_{ab} = \mathbf{d}\cdot\mathbf{e}_x$, $f(x)$ is the wave function of the antenna quasiparticle. The normally ordered observable value of the intensity in the Fraunhofer zone is given by

$$\langle \hat{\mathbf{E}}^{(-)}(\mathbf{r},t)\cdot\hat{\mathbf{E}}^{(+)}(\mathbf{r},t)\rangle = \frac{\omega_0^4 d_{ab}^2}{(4\pi\varepsilon_0)^2 c^4} \int_{-l/2}^{l/2}\int_{-l/2}^{l/2} \frac{f(x)}{|\mathbf{r}-\mathbf{r}'(x)|} \frac{f^*(x')}{|\mathbf{r}-\mathbf{r}'(x')|} K(x,x';t) dx' dx \quad (16)$$

where

$$K(x,x';t) = \left\langle \hat{\sigma}^+\left(t - \frac{|\mathbf{r}-\mathbf{r}'(x)|}{c}\right) \cdot \hat{\sigma}^-\left(t - \frac{|\mathbf{r}-\mathbf{r}'(x')|}{c}\right) \right\rangle$$

(17)

For the steady state of the antenna produced by coherent external driving, the correlation function (16) is independent of time. Equation (17) can be rewritten as

$$K(x,x') = \left\langle \hat{\sigma}^+(0)\cdot\hat{\sigma}^-\left(\frac{|\mathbf{r}-\mathbf{r}'(x)|-|\mathbf{r}-\mathbf{r}'(x')|}{c}\right) \right\rangle \quad (18)$$

The time shift in the case of the antenna is stipulated by the phase shift of the radiation from different points of the antenna and is given by $\tau = (|\mathbf{r}-\mathbf{r}'(x)|-|\mathbf{r}-\mathbf{r}'(x')|)/c$. For simplifying the integration in Equation (17), we use the conventional assumptions of the macroscopic antenna theory [5]:

$$|\mathbf{r}-\mathbf{r}'(x')| \approx R \quad (19)$$

$$|\mathbf{r}-\mathbf{r}'(x')| \approx R - x'\cos\theta, \qquad (20)$$

for the amplitude and phase factors, respectively, where $R, \theta$ are the coordinates for the spherical system with the origin at the antenna's center. As a result, the intensity of the radiation can be presented in terms of the radiation pattern, conventional for the antenna theory. It reads

$$\langle \hat{\mathbf{E}}^{(-)}(\mathbf{r},t) \cdot \hat{\mathbf{E}}^{(+)}(\mathbf{r},t) \rangle \approx \frac{\omega_0^4 d_{ab}^2}{(4\pi\varepsilon_0)^2 c^4} \frac{1}{R^2} \cos^2\theta \cdot \xi(\theta) \qquad (21)$$

where

$$\xi(\theta) = 2\operatorname{Re} \int_{-l/2}^{l/2} dx \cdot f(x) \int_x^{l/2} f^*(x') \left\langle \hat{\sigma}^+(0)\hat{\sigma}^-\left(\frac{x'-x}{c}\cos\theta\right)\right\rangle dx' \qquad (22)$$

($\theta < |\pi/2|$). Equations (21, 22) describe the energy radiation pattern of a quantum dipole antenna. The factor (22) shows the difference between an elementary Hertz dipole and a dipole antenna in the classical case.

Finally, in the end of this subsection, we address the problem of feeding our single TLS quantum antenna. Generally, by simply connecting it to another quantum system, a compound antenna is just formed. For example, if the feed is taken as another TLS, described by the creation and annihilation operators $\hat{s}^{\pm}$ and linearly coupled to the original TLS, $\hat{V}_{feed-emitter} = \hbar\Omega(\hat{s}^+\hat{\sigma}^- + \hat{\sigma}^+\hat{s}^-)$, one has the oscillatory exchange of excitations between the TLSs. The target of the feed will back-act on the feed. However, it is still possible to find quantum systems practically unaffected by the interaction with the antenna. For example, one can shine an intense resonant coherent beam on the TLS having $\hat{V}_{feed-emitter} \approx \hbar\Omega(t)(\hat{\sigma}^- + \hat{\sigma}^+)$, where the Rabi frequency $\Omega(t)$ describes the product of the coherent state amplitude and the interaction constant. In that case, the state of the beam is practically unaffected by short interaction with the TLS [54]. The so-called $\pi$-pulse (i.e. a rectangular pulse satisfying $=\pi$) completely transfers the TLS population from the ground state to the excited state. Another way of feeding the TLS antenna is to connect it with a large classical object such as an additional thermal reservoir providing the so-called "incoherent pump" [38].

Notice that there is a way of limiting the back-action of the antenna on the quantum feed by implementing the so-called quantum cascade scheme [55–57]. The feed (it can even be a small set of quantum systems, such as a mode or a Jaynes-Cummings system [58–60]) couples to the antenna through the common dissipative reservoir damping the back-action. However, to prevent the back-action completely, one would still need something akin to topological structures supporting the field propagation only in one direction [61].

### 2.2. Two-atom antenna

While highlighting specifically quantum features of the quantum antenna set-up, the single-atom antenna lacks the important property of a multi-emitter arrangement: the possibility to combine the effects of the classical interference, appearing due to different placement of the emitters, and the quantum interference due to the quantum correlations of different components of a quantum antenna. Also, a number of the effects arises due to coupling of the emitters to the same reservoir of the electromagnetic field modes. To highlight these effects, let us consider a simple generalization of a single TLS antenna: two TLS emitters placed in the homogeneous space. In the Markovian approximation, one can obtain analytically the solutions for both antenna variables, and for the field in the far-field zone[24].

The master equation, describing two interacting identical TLS in the Markovian field reservoir and in the basis rotating with the TLS transition frequency, can be written in the following form[24]

$$\frac{d}{dt}\hat{\rho} = i[\hat{V}_2, \hat{\rho}] + \frac{1}{2}\sum_{j,l=1}^{2} \gamma_{jl} \left(2\hat{\sigma}_j^- \hat{\rho}\hat{\sigma}_l^+ - \hat{\rho}\hat{\sigma}_l^+ \hat{\sigma}_j^- - \hat{\sigma}_l^+ \hat{\sigma}_j^- \hat{\rho}\right), \qquad (23)$$

where

$$\hat{V}_2 = f_{12}\hat{\sigma}_1^+ \hat{\sigma}_2^- + h.c. \qquad (24)$$

The remarkable fact about Eqs, (23,24) is the appearance of non-unitary, so-called "dissipative" coupling through the reservoir described by the rate $\gamma_{12} = \gamma_{21}$ in addition to the unitary coupling (24). The rate of the unitary coupling, $f_{12}$, tends to the infinity when the distance between TLS tends to zero. However, at the same time, the dissipative coupling rate $\gamma_{12}$ remains finite and tends to the single-TLS population decay rate, $\gamma_{11} = \gamma_{22}$, defined by the "Fermi's golden rule"(see Reference [40] ). Also, one always has $\gamma_{11} \geq |\gamma_{12}|$.

The time-dependent population of each TLS has two components: the first one decaying faster than the single TLS population with the rate $\gamma_{11} + |\gamma_{12}|$ (so-called "super-radiant" component), and the other one decaying slower than the single TLS population with the $\gamma_{11} - |\gamma_{12}|$ (so-called "sub-radiant" component). These components are closely connected with the symmetry of possible two-TLS states. For the close TLS, when $\gamma_{11} \approx \gamma_{12}$, the anti-symmetric state $|+>_1 |->_2 - |+>_2 |->_1$ becomes the "dark state". It is not affected by losses, whereas the population corresponding to the symmetric state $|+>_1 |->_2 + |+>_2 |->_1$ decays with twice the rate of the single TLS. Without the unitary exchange, the latter case would correspond to the famous Dicke model of superradiance[62]. Notice that this model is quite difficult to realize in experiment. Only recently in the set-up somewhat close to the discussed two-TLS scheme (namely, two superconducting qubits coupled to a leaky microwave cavity), the supperradiant decay has been observed experimentally[63].

For an antenna-like arrangement with the TLS separated by the distance of the order of the wavelength, both super and sub-radiant components are essentially present. In the process of decay even two completely initially excited (and thus uncorrelated) TLS, the entanglement arises between the TLS[64]. The quantum correlations are transferred to the emitted field. Thus, the correlation functions of the higher order cannot be represented through the correlation functions of the lower order, as it is for the classical emitters producing coherent states. The two-TLS antenna exhibits rather a complicated pattern of spatial and temporal correlations in the far-field.

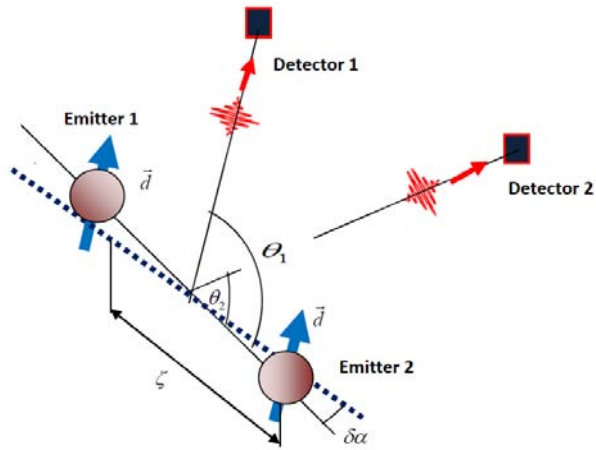

**Figure 3**. A schematic illustration of the TLS quantum antenna dimer, and the detection of the emitted photon pairs.

Let us consider the detection of a photon pair emitted by our antenna as depicted in the scheme shown in **Figure 3**. For the sake of simplicity, we limit ourselves only with the plane perpendicular to the dipole moments. The probability of detecting the first photon in a small solid angle at the direction $\theta_1$ and the time $t$ and the second photon at the direction $\theta_2$ and the time $t + \tau$ is proportional to the second-order intensity correlation function of the field[24]:

$$\langle \hat{\mathbf{E}}^{(-)}(\theta_1, t)\hat{\mathbf{E}}^{(-)}(\theta_2, t+\tau)\,\hat{\mathbf{E}}^{(+)}(\theta_2, t+\tau)\hat{\mathbf{E}}^{(-)}(\theta_1, t)\rangle \propto$$
$$exp\{-\gamma_{11}\tau\}[(1 + \cos\varphi(\theta_1)\cos\varphi(\theta_2))\cosh\gamma_{12}\tau - (\sin\varphi(\theta_1) + \sin\varphi(\theta_2))\sinh\gamma_{12}\tau +$$
$$\sin\varphi(\theta_1)\sin\varphi(\theta_2)\,\cos f_{12}\tau\,], \qquad (25)$$

where the function $\varphi(\theta) = \frac{\varsigma\omega}{c}\cos\theta$, $\varsigma$ is the distance between the dipoles.

The spatial distribution of the probability of the photon pair registration given by Equation (25) is rather strongly dependent on the delay. For $\tau = 0$, the spatial pattern is the same as for a non-interacting TLS (i.e. for $\gamma_{12}, f_{12} = 0$ ): $\langle \hat{\mathbf{E}}^{(-)}(\theta_1, t)\hat{\mathbf{E}}^{(-)}(\theta_2, t)\,\hat{\mathbf{E}}^{(+)}(\theta_2, t)\hat{\mathbf{E}}^{(-)}(\theta_1, t)\rangle \propto [1 + \cos(\varphi(\theta_1) - \varphi(\theta_2))]$. Only the effects of the classical interference of two differently placed dipole emitters show up for the case.

Another curious similarity with a classical two-dipole antenna is the formal equivalence of the equations for the TLS and field averages obtained from Equations (24,25) and the semiclassical Maxwell-Bloch equation, describing dipole dynamics affected by each other emitted fields[65]. Assuming the factorization of the field and TLS averages, one can exactly obtain one system of equations from the other. Adding noise, it is possible even to reproduce super-radiance effect with semi-classical equations[65]. However, the spatial profiles of the higher-order correlation functions in semiclassical and quantum (i.e. without the averages factorization) cases remain drastically different[66].

Notice that feeding the two-TLS antenna is somewhat more difficult than the single TLS. The reason for that is the so-called "dipole blockade" arising due to the interaction of the TLS [67]. Essentially, the absorption of the first photon by the first TLS induces an interaction-dependent frequency shift in the second TLS lowering the probability of the excitation by the semiclassical field resonant with the unperturbed TLS transition. For sufficiently strong interaction between the TLS, this effect can effectively limit the excitation of the antenna to single-excitation subspace for an insufficiently intensive pumping [68]. Notice that this interaction can be dissipative as well [69].

From the other side, the interplay between the quantum and classical interference and the essential non-linearity of the TLS antenna, allows one to control the emission of the antenna by implementing a field structured on the antenna size scale to drive it. For example, using Laguerre-Gaussian or Bessel beams for a pump, one can achieve highly tunable directional emission [70].

### 2.3. Multi-atomic antenna: approaches to description

When trying to consider more complicated emitters arrangements, one immediately encounters the problem typical for any multi-particle quantum system: rapid growth of the number of variables, necessary to describe the state of the system, with the size of that system. For example, for $M$ TLS, one needs to find $4^M$ components of the density matrix. So, in practice, one can find an exact solution at most for a few tens of TLS. For example, as it is bluntly stated in the recent work on superradiance [71], for 200 emitters an exact solution "is not feasible due to the huge state space".

For large quantum antennas (i.e. for $M \gg 1$), there is an important case of low antenna excitation, when the maximal number of excitations in the antenna is much less than $M$. For a few excitations, one can even derive an exact solution, and, for a larger number of excitations, one can recourse to the Holstein-Primakoff approximation [72]. To see the essence of it, let us write the multi-TLS generalization of the interaction Hamiltonian (11),

$$\hat{V}_{emitter-field} = \hbar \sum_{\forall k,j}(v_{kj}\hat{\sigma}_j^+ \hat{a}_k + h.c.), \qquad (26)$$

where the constants $v_{kj}$ describe the interaction of the $j$-th TLS with the $k$-th mode of the field. Different TLS are independent, so, one has

$$\langle[\hat{\sigma}_j^-, \hat{\sigma}_l^+]\rangle = \delta_{jl}\left(1 - 2n_j^{(+)}\right), \qquad (27)$$

where $n_j^{(+)}$ is the upper-level population of the $j$-th TLS. If this population is much less than unity, one can replace the TLS rising and lowering operators $\hat{\sigma}_j^\pm$ with the bosonic creation and annihilation operators, and linearize the problem. This simple approach was successfully applied, for example, for the description of the entangled state generation by complex multi-TLS antennas, for taking into account the emitters movement [73], for the description of electromagnetically induced transparency[74], and controllable sub- or superradiant decay in an ensemble of weakly driven emitters[75].

One can step beyond the condition of the low excitation (27) with the Holstein-Primakoff approximation, if the TLS are identical and placed closely together, so the constants $v_{kj}$ are the same for all $k$ and $j$. Then, it is possible to introduce collective operators

$$\hat{S}^{\pm} \propto \sum_{\forall j} \hat{\sigma}_j^{\pm} \tag{28}$$

and replace them with the bosonic operators $\hat{A}^+, \hat{A}$:

$$\hat{S}^- \to \sqrt{M - \hat{A}^+\hat{A}}\,\hat{A}; \quad \hat{S}^+ \to \hat{A}^+\sqrt{M - \hat{A}^+\hat{A}} \tag{29}$$

for the number of excitations still much less than M. It should be noted that the nonlinear transformation (29) can be applied even to replace the single TLS operators with the bosonic ones. Furthermore, there is a possibility to generalize the collective operators (28) for the case of non-homogeneous coupling to the field [76].

Notice that the collectivity of the emitters allows one to greatly simplify the description of the dynamics using the collective states produced by the consequent action of the operator $\hat{S}^+$ on the lowest energy states of all the emitters. These states are called "Dicke states" [77]. They are permutationally invariant with respect to the individual emitters, entangled apart from the lowest and the highest energy states, and there exist only $M+1$ of them for $M$ TLS [78]. The concept of the Dicke states can be generalized for a set of independent emitters subject to the same Markovian loss[78,79].

Another approach potentially working for quite large ensembles of emitters is to derive the equations for operators averages and assume that the correlations between them are small enough to eventually applye the approximation $\langle XY \rangle \approx \langle X \rangle \langle Y \rangle$. This makes the equation nonlinear, reduces the number of variables and closes the system of equations (see, for example, the cumulant expansion technique[38]).

## 2.4. Multi-atomic antenna: no "talking" between emitters

An important simple case of the quantum antenna is when there is no "talking", i.e. interaction between the emitters. Each emitter emits individually as it was described in Subsection 2.1. However, the initial state of the antenna can be non-classical and correlated, and the emitted field is affected by both the quantum and classical interference leading to a number of quite non-trivial effects.

The first and arguably the most known of them is the effect of directional emission, stemming from the excitation of an ensemble of randomly positioned emitters in a specific entangled "timed Dicke state" [80]. If one illuminated the ensemble of randomly placed position-fixed emitters with a plane wave carrying just one photon, and this photon were absorbed, the emitted photon would fly exactly in the same direction, and the emitted field would be the plane wave identical to the impinging one. This effect occurs due to the mapping of the impinging state to the emitters' ensemble acting like perfect quantum memory.

As the result of the photon absorption, the following state with the entangled *M*-emitters ensemble is created

$$|\Psi\rangle = \frac{1}{\sqrt{M}} \sum_{j=1}^{M} \exp\{-i\mathbf{k}\mathbf{r}_j\} |+\rangle_j \quad , \tag{30}$$

where **k** is the wave-vector of the plane wave, the number of the emitters *M* >>*1*, the vectors |+>*j* are the states with the *j*-th emitter in the excited state, all the other emitters are in the ground states, and **r**$_j$ is the position of the *j*-th emitter. The state (30) was named a "timed Dicke state" for being similar to the usual single-excitation Dicke state with the same probabilities to find every emitter in the excited state. However, the phases in Equation (30) carry the information about the impinging state (more precisely, about the moment of time when the impinging wave reached a particular emitter). If one reverses the signs of the phases in Equation (30), the emitted photon would fly in the opposite direction.

Regular arrangements of emitters (i.d. similar to conventional antennas) can also exhibit a number of effects produced by the interplay of the quantum and classical interference effects. For example, emitters arrays in Dicke states can manifest superradiant peaking of the emitted field intensity, when the maximal intensity of the field is proportional to the square of the number of the excited emitters [81]. Also, the spatial distribution of the emitter field can be either strongly narrowed or spread by using different kinds of entangled states of the antenna. Moreover, the emitted light can exhibit strong superbunching [82].

Notice that quantum correlations between the emitters can be designed to produce the desired spatial distribution of not only the emitted field intensity, but also the field correlation functions of a higher order. Thus, one can generalize the classical concept of directivity for the correlation functions requiring maximization for certain observation directions [83,84]. Curiously, the photon number statistics of the registered field also becomes direction-dependent. Here, one can notice that the classical interference of incoherent uncorrelated emitters can also exhibit supperradiance by specifically arranged detection of *N*-order correlation functions: to that end, one photon should be detected at a certain angle and all the other *N-1* photons at other angle [85].

The state of an antenna can be designed to provide the desired correlations between the emitted photons. It can be achieved, for example, by finding the state, maximizing the generalized directivity. In the work [66], it was shown how to design an array antenna state to obtain co- or contra-directionally correlated photons, or even completely suppress the field in the far-field zone realizing quantum analogy of a "non-radiating source" [86].

**2.5. Multi-atomic antenna: single-photon case**

As was mentioned in Section 2.3, it is quite hard to realize the "original" superradiant Dicke model. When one brings the emitters closely together, they start interacting in both unitary and dissipative ways. So, to observe "clean" Dicke-like enhancement of the radiation rate, one generally needs to strongly modify the environment to couple the emitters to the field in the same way, and simultaneously place them far enough from each other to prevent them from interacting, using, for example, a single-mode leaky cavity/waveguide [63,87]. As it was also mentioned in Section 2.3, the description of the multi-excitation dynamics in the multi-emitter antenna is a formidable task. Nevertheless, a lot of features of collective behavior (including Dicke superradiance) can be captured with a single-excitation problem, when the antenna emits just a single photon.

Formally, for that case, the problem of emission coincides with that of the emission by an antenna with harmonic oscillator emitters [88]. It can be solved just like a single TLS-problem described by the Hamiltonian (10), and the solution can be represented in the form similar to Equation (12):

$$|\psi(t)> = \sum_{m=1}^{M} c_{0m}(t) |+>_m |vacuum> + \sum_{\forall k} c_k(t) |-> |1>_k, \qquad (31)$$

where the vector $|->$ denotes the ground state of all the M emitters, the state $|+>_m$ corresponds to the $m$-th exited emitter, and all the other are in the ground states. Under the Markovian approximation and for simple field reservoirs (such as, for example, homogeneous vacuum or a leaky cavity), the coefficients $c_x(t)$ can be found analytically, the collective Lamb shift and superradiant decay rates can be easily estimated. This fact allows one even to design the antenna arrays for tailoring superradiance [89]. The validity of the single-photon superradiance model was confirmed in the experiments with cold atoms, and linear scaling with the number of atoms was confirmed as well [90].

Notice that in a multi-emitter antenna not only superradiant, but also sub-radiant effects can strongly manifest themselves. For a larger number of emitters, the presence of sub-radiant states can lead to very slowly decaying correlations between the emitters and the field. For example, in the recent work [91], the decay rate of 100 times less than the single-emitters decay rate was measured. Even in a 1D antenna array, one can realistically design the single-excitation states with more than $10^5$ times slower decay rate for 100 emitters [92]. Thus, for such cases, one should be rather careful about the application of the Markovian approximation. Subradiance can be potentially used for storing quantum states in an antenna and a fast readout of this state by switching the antenna into a super-radiant regime [93,94].

Arranging the emitters, one can tailor a spatial and temporal form of the emitted single-photon pulse [95–97]. Notice that in analogy with the "timed Dicke states" emission, higher directivity of the emitted field intensity is reached for more homogeneous emitters distribution [96]. It should be noted that not only superradiant, but also subradiant emitted field can be shaped [97,98].

Curiously, one can reach quite high directivity and spatial pulse shaping with rather a modest number of quantum emitters by implementing a structure with a chiral light-atom interface [99], where the absorption and the spontaneous emission of light are essentially unidirectional [100].

Even with single-excitation states (in particular, the timed Dicke states) of the quantum antennas and a single linear array structure, they can exhibit breaking of the reciprocity theorem [101]. Let us assume that the near field has the form of a travelling wave, $e^{ij\beta}$, with an arbitrary phase shift $\beta$. The radiation pattern is invariant under the exchange $F(\theta,\omega,\beta,\mathbf{k}) = F(\pi-\theta,\omega,-\beta,-\mathbf{k})$, but not $F(\theta,\omega,\beta,\mathbf{k}) = F(\pi-\theta,\omega,-\beta,\mathbf{k})$ ($\theta$ is the meridian spherical angle). This situation is similar to the Onsager relation for kinetic coefficients for a uniformly rotating body and for bodies in the external magnetic field (in this case, the vector $\mathbf{k}$ plays the role of rotation velocity or induction of the magnetic field).

The corresponding operator of the array field in the Fraunhofer zone is

$$\hat{\mathbf{E}}^{(+)}(\mathbf{r},t) = -\mathbf{e}_\theta \frac{\omega^2 d_{ab}}{4\pi\varepsilon_0 c^2 R} \sin\theta \, \hat{\sigma}^-_{Far}\left(t - \frac{R}{c}, \theta\right) \qquad (32)$$

where $\hat{\sigma}^-_{Far}(t-R/c,\theta) = \sum_{j=1}^M \hat{\sigma}^-_j(t - jac^{-1}\sin\theta)$. The near field reads

$$\hat{\mathbf{E}}^+_{Near} \big|_{\ldots} = \sum_{j=1}^M \hat{\sigma}^-_j(t) \qquad (33)$$

The antenna emission can be described as a transformation of the near field to the far field

$$\hat{\sigma}^-_{Far}(t,\theta) = \hat{U}^{-1}(\theta) \hat{\sigma}^-_{Near}(t) \hat{U}(\theta), \qquad (34)$$

where $\hat{U}(\theta) = \prod_{j=1}^M \hat{U}_j(jac^{-1}\sin\theta)$ and $\hat{U}_j(\tau) = \begin{pmatrix} e^{-i\omega\tau/2} & 0 \\ 0 & e^{i\omega\tau/2} \end{pmatrix}_j$ is the partial operator acting on the states of the atom with the number $j$[102]. Equation (34) has a conventional form of an evolution equation in the Heisenberg picture[2], but the field evolves with the spherical angle on the equiphase surface instead of the evolution with time. As it was shown in Reference[102], the near and far fields, given by Equations (32) and (33), respectively, are non-commutative. It leads to the manifestation of the Heisenberg uncertainty principle [103,104] for a 1-D antenna array of the two-level quantum emitters. The far-zone uncertainty for timed Dicke state given by Equation (30) exhibits a multi-lobe form consisting of the major and minor lobes alternating with the uncertainty free ("dark") angular directions. The uncertainty pattern looks similar to the radiation pattern of the antenna but does not follow it identically. As a result, the quantum antenna emission can be accompanied by the directive squeezing of radiation; the strong values of squeezing are reachable in the narrow directions of high emission. The squeezing mechanism is not similar to the well-known degenerate parametric amplification[2]. Squeezing in quantum antennas is a promising tool for the noise suppression in the quantum informatics and quantum measurements [102], as well as for its primary application in the interferometry[105].

Finally, we note that the case of a multi-level system with just a single excitation is methodologically similar to the problem considered in this Subsection. Such a system can also exhibit specifically quantum effects. In particular, it has recently been shown that a system with one upper and three degenerate lower levels can act as an isotropic unpolarized coherent single-photon emitter[45].

## 2.6. Multi-atomic antenna: multi-photon case

A few excitations in the interacting ensemble of emitters of the quantum antenna can still be tractable. For example, the master equation similar to (23) for $M$ TLS of the "talking" quantum antenna

$$\frac{d}{dt}\hat{\rho} = i\left[\hat{V}_M, \hat{\rho}\right] + \frac{1}{2}\sum_{j,l=1}^{M} \gamma_{jl}\left(2\hat{\sigma}_j^-\hat{\rho}\hat{\sigma}_l^+ - \hat{\rho}\hat{\sigma}_l^+\hat{\sigma}_j^- - \hat{\sigma}_l^+\hat{\sigma}_j^-\hat{\rho}\right) \quad (35)$$

for a few excitations can be solved directly, and, therefore, the field in the far-field zone can be explicitly found (by summing the single-emitter terms as it is in Equation (24)).

The main difference from the single-excitation case considered in the previous subsection is the possibility to capture deviations from linear (harmonic-oscillator like) dynamics caused by a finite number of emitter levels (and also not captured by the Holstein-Primakoff approximation). The correlations between the antenna excitations are translated into the correlations between the emitted photons. This enables, for example, the generation of the two-photon states that are entangled in orbital angular momentum by a symmetrically excited ring antenna [106], to reveal the presence of localized bound states in the antenna [107] and shape spatial photon correlations for the emitting subradiant states [108].

Notice that one can close the system of equations stemming from Equation (35) by assuming the quantum correlations between the emitters to be weak, in order to express the averages of the products of any $L$ operators through the products of the averages of less than $L$ operators. This approach allows one to capture some features of the emission caused by the emitters correlation through the common field reservoir. For example, the supperradiance from multi-level emitters can be successfully described in this way taking account of two- and more photon correlation [109–111]. Also, one can capture the correlation features of the field up to a certain order. It is comparatively simple, for example, to model a second-order correlation function of a superradiant pulse. In this way, it was shown that the supperradiant pulse can carry a strongly superbunched state [71]. Curiously, the maximal superbunching occurs at the edges of the superradiant pulse, and is maximal for a few emitters degrading with an increas of their number [71].

Notice that both superradiant and subradiant emission can occur not only in antennas, initially prepared in specific states, but also in coherently [112] and even incoherently driven emitters systems [113].

Here, one should mention another interesting effect, occurring in a system of non-identical emitters (i.e. detuned one from another, with different decay rates). This is emitters synchronization, arising due to the nonlinearity of the interaction between them [114]. Despite the irregularities in location and parameters of the emitters and incoherent pumping, a steady state can appear that exhibits macroscopic quantum-phase coherence and quantum correlations between the emitters [113]. Such a state is usually accompanied by subradiant emission [115].

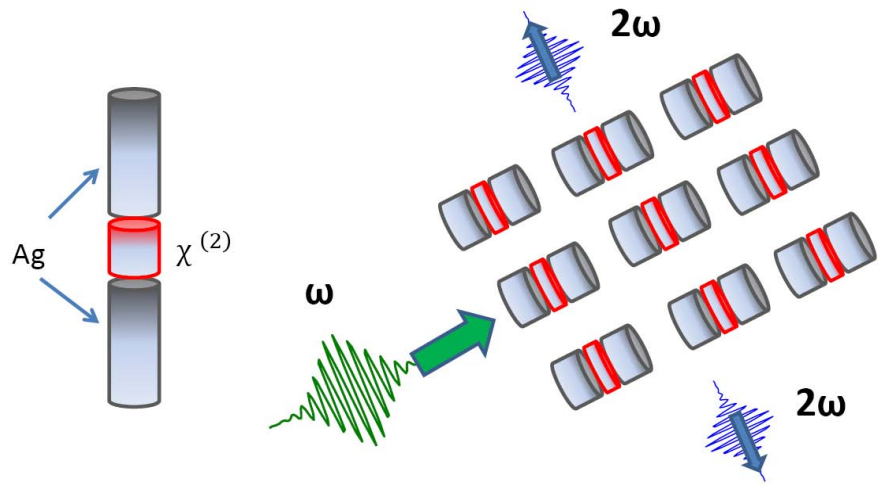

**Figure 4.** Design and operation principle of a nanoscale second-harmonic frequency conversion system. It consists of a plasmonic metasurface made of an array of dipole nanoantennas loaded with $\chi^{(2)}$ nonlinear optical materials [116].

## 3. NONLINEAR ANTENNAS

### 3.1. Antennas with weak nonlinearity

Initially, antennas were almost exclusively passive linear devices loaded by independent lumped nonlinear elements (such as, for example, semiconductor diodes and transistors). Toward the end of 20$^{th}$ century, there was a tendency to transfer the optical concepts and arrangements to the low frequency ranges (mm and sub-mm wavelengths). A typical example was the usage of Fabry-Perot interferometers instead of conventional closed single-mode resonant cavities. The progress in nanotechnologies led to the inversion of this tendency: the radio-frequency concepts (antennas, networks, transmission lines) are translated to the optical domain. In particular, the nonlinear nanostructures of different types are considered and arranged as nanoantennas[43,116–118]. The nanoantennas are actually used as the bridge between nanoscale signal processing and free-space radiation. In spite of a small volume of nonlinear material load and its naturally weak response, the nanoantenna geometry can significantly enhance the nonlinear optical properties of a device for rather low light intensities.

      Nonlinear response of the nanoantennas strongly depends on their configuration, local environment and the polarization. One of the simplest types of a nonlinear nanoantenna consists of a horizontal periodic structure of elementary dipole emitters (see **Figure 4**). An elementary emitter consists of silver nanowires separated by a small gap loaded with χ (2) nonlinear optical materials. The nonlinear frequency mixing is induced at each nanoload. The period of the nanowires lattice is much smaller than the wavelength of the pumping field. This implies that it can be possible to implement dense arrays of resonant nanoantennas, which may be considered as effective artificial nonlinear metasurfaces[116]. They open up the promising opportunities for the second-harmonic generation, sum/difference frequency generation, surface-enhanced Raman scattering and optical bistability[116].

As suitable materials for nanoloads, III – V compounds have been studied, such as GaAs and CdGeAs$_2$ with the typical values of linear permittivity $\varepsilon_L = 12.5$ and the second-order susceptibility $\chi^{(2)} \cong 250$ pm/V, respectively [119]. This nonlinear metasurface can assist various applications based on the second-order harmonic effects, paving the way to near-field imaging and spectroscopy, chemical sensing, localized photon sources, etc. The observed sharp peak of the converted second-order harmonic efficiency is 0.0034 % [116].

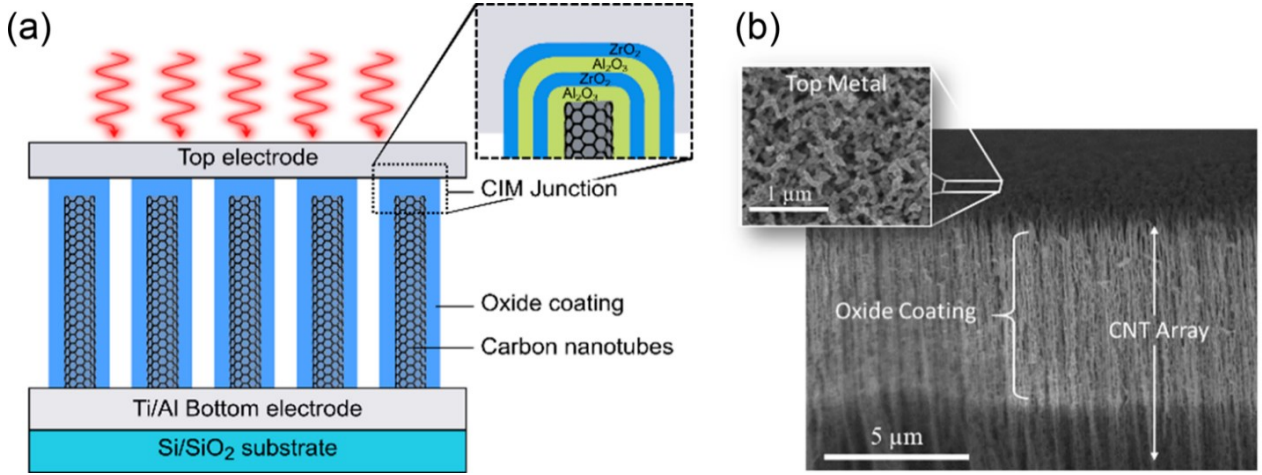

**Figure 5.** (a) CNT rectenna device. (Inset) The tips of the CNTs are coated in a quad-insulator laminate of dielectric and capped with an Al top metal electrode to form a CNT/quad-insulator/metal (CI$^4$M) tunneling diode. (b) SEM side view of the CNT array coated with an oxide and top metal. The oxide penetrates several microns into the array while conformally coating the CNTs. (Inset) The top metal, coating the array, is not planar but rather forms an interdigitated network of metal-coated CNTs with the gaps between them. Reprinted with the permission from ACS Appl. Electron. Mater. **2019**, 1, 5, 692[123]. Copyright (2019) American Chemical Society

Another set of the proposed applications is based on the usage of plasmonic metasurfaces loaded with $\chi^{(3)}$ materials. For example, as it was mentioned in Reference [116], the non-linearity for fused silica has the Kerr form $\varepsilon_{load} = \varepsilon_L + \chi^{(3)} |E|^2$ with $\varepsilon_L = 2.2$ and $\chi^{(3)} = 4.4 \times 10^{-20}$ m$^2$/V$^2$. These typical values make possible the antenna-load matching and tuning. Such types of the quantum antennas open the ways for enhancement of optical bistability and optical switching. Another way is the usage of intrinsic nonlinearities of plasmonic metals for the strong local field enhancement at the surface and sharp edges of nanoantennas (for example, as it was reported in Reference [116], $\chi^{(3)} \cong 10^{-9}$ m$^2$/V$^2$).

The remarkable electronic properties of carbon-based materials (graphene, single-wall CNTs, multi-wall CNTs) also open the way for their usage in non-linear nanoantennas. The general concepts of dipole antennas are similar to those for plasmonic metals. The first one is loading CNTs by lumped non-linear elements. The second one is the usage of the intrinsic nonlinearities of CNTs. In such a way, the vertical and horizontal configurations of antenna arrays have been implemented. The horizontal configuration is considered in [120–122]. The example of the vertically aligned array of multi-walled CNTs is shown in **Figure 5**[123].

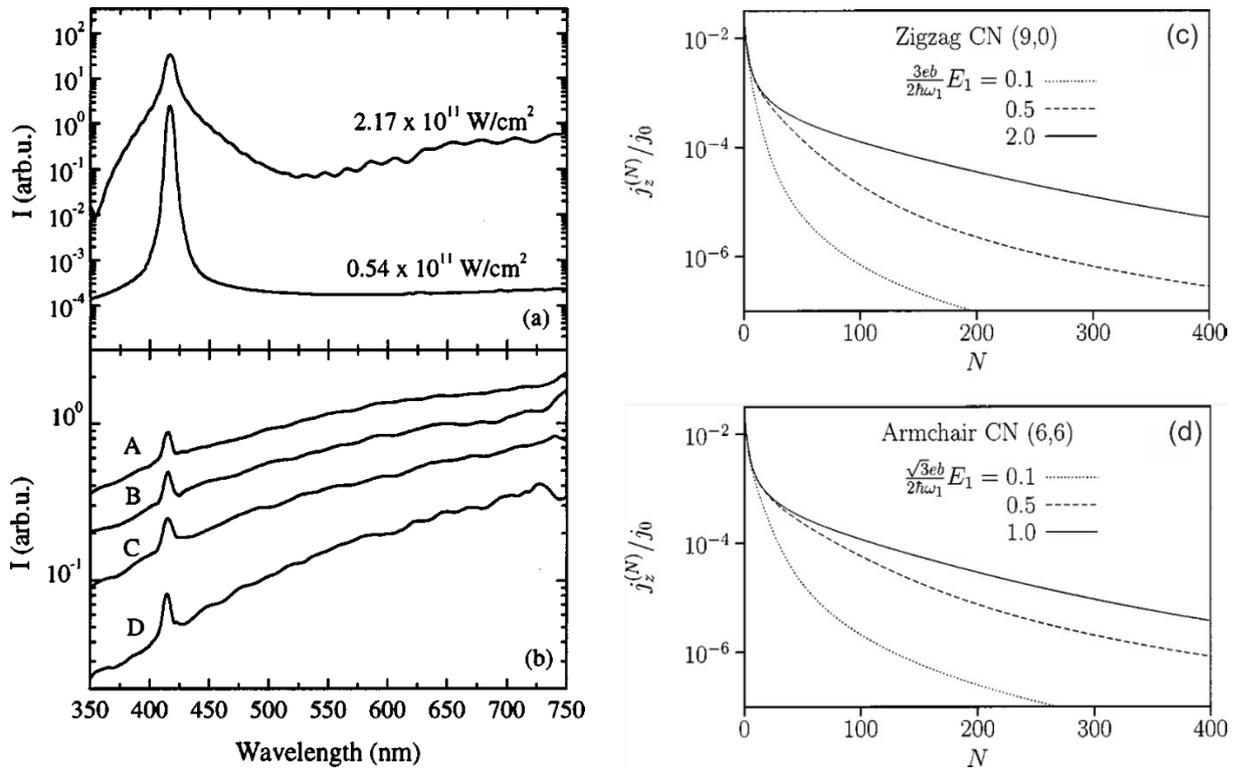

**Figure 6**. Nonlinear optical properties of CNTs [120–122]. a),b) Broad background and 3-rd harmonic signal generated by the interaction of the intense laser radiation with the aligned MWNTs: a) theory; b) experiment. The input intensities: curve A - $2.3\times10^{11}$ W/cm$^2$, curve B - $1.7\times10^{11}$ W/cm$^2$, curve C - $1.3\times10^{11}$ W/cm$^2$, curve D - $0.8\times10^{11}$ W/cm$^2$. c),d) Theoretical high-order harmonic spectra of the nonlinear current induced in the metallic zigzag and armchair CNs by the different driving field strengths as indicated. The normalization is $j_0 = 5.5\times10^6$ A/m for c) and $j_0 = 4.6\times10^6$ A/m for d). Reprinted with the permission from References[120,121]. Copyright (2019) American Physical Society.

One of the promising areas of the non-linear CNT-antennas is the emision of high-order harmonics. The experimental results for the third harmonic generation from solid samples of CNTs are given in Reference[122]. To investigate the effect of alignment on the measurements, the used samples consisted of self-standing MWNTs perpendicular on Si(001) substrates, which were deposited by thermal chemical vapor deposition. The results are interpreted on the basis of the full quantum mechanical theory for the harmonics generation from a single-walled CNT. **Figures 6a,b** show the intensity of light, generated in the spectral range 300–750 nm around the 3-rd harmonic of the Cr:Forsterite laser at 417 nm. They demonstrate that these spectra represent continuous background superimposed by a narrow spectral line corresponding to the 3-rd harmonic of the pump frequency. The high value of the intensity occurs for the fifth harmonic, too [122]. As it was measured in Reference [122], the third-order susceptibility for CNT-metamaterial is estimated as $\left|\chi^{(3)}\right| \cong 10^{-6}\,\text{m}^2/\text{V}^2$, which exceeds by three orders the corresponding value for plasmonic metals.

As it was shown in References [120,121], the mechanism of the nonlinearity, provided by the motion of $\pi-$ electrons in metallic CNTs, results in an efficient generation of high-order harmonics with the intensities that fall down smoothly with a harmonic number, without a cutoff. The qualitative behavior of the armchair CNT spectrum is shown in **Figures 6c,d**. The high-level intensities of the harmonics with the orders $N$=100-150 open the way for their practical applications.

### 3.2. Strong coupling: Rabi oscillations, Rabi-Bloch oscillations and Rabi waves

In this Subsection, we will focus on the principles of electrical tuning of the antenna parameters, made possible by strong coupling of the antenna with the electromagnetic field. We will consider two models of quantum antennas in a strong coupling regime. The first one is a loop, which is equal to an elementary magnetic dipole [5]. The second one is an electrical dipole antenna corresponding to a single emitter placed at a thin nanowire. The theory, described in this Subsection, is developed in References [31,124] and is based on the theory of the open quantum systems. For example, in Reference [31], the open-system approach is applied to the description of the tunnel coupling between the neighboring emitters in the periodic loop of two-level emitters. The given emitter is considered as a quantum system, which is described by the reduced density matrix. The neighboring emitters form a reservoir with respect to the given one, and the electron tunneling over the barrier is considered as the "effective losses". In the model [124], the separated antenna is considered as a quantum system, while the surrounding space is considered as a photonic reservoir. The transformation of the near field to the far field is described as the "effective losses" of the antenna energy.

Let us consider in more details the example of an infinite periodical two-dimensional (2D) square lattice of identical two-level QDs, with the distance between the dots being $a$ [125]. The interlevel transition in the QD is described similarly to the model given in **Section** 2. The neighboring QDs are coupled through the electron tunneling, with the tunneling frequencies being $\xi_{1,2}$ for the ground and the excited states, correspondingly. Only intraband tunneling transitions are permitted. The quantities $\xi_{1,2}$ and the dipole moments **d** are the phenomenological parameters, whose realistic values have been estimated in Reference [30]. Let the QD array be exposed to classical light, $\mathbf{E}(\mathbf{r})=\mathbf{E}_0(\mathbf{r})\exp(i(\mathbf{kr}-\omega t))$, where **k** and ω are the frequency and the wave vector of the operating field, and **r** lies in the plane (x,y). The Hamiltonian of the QD array is $\hat{H}=\hat{H}_0+\hat{H}_T$, where the term $\hat{H}_0$ describes the system in the absence of electron tunneling. The term $\hat{H}_T$ accounts for the electron tunneling between the dots and has the form

$$\hat{H}_T = -\hbar \sum_{pq} \left( \xi_{pq} \cdots \right) \hbar \sum_{pq} \left( \xi_{p,q} \cdots \right)_{pq} \cdot_{p-1,q-1} \langle -| \right) \quad (36)$$
,

where the particular QDs in the lattice are labeled with the indexes $p,q$. This Hamiltonian can be considered as a component responsible for spatial propagation of Rabi waves in the QD array [30]. The problem is simplified via the transformation to the spatially continuous limit ($p \to \mathbf{r}, q \to \mathbf{r}'$). The dynamics of a single-particle state in the QD-array is described by the following effective density matrix:

$$\hat{\rho}(\mathbf{r},\mathbf{r}',t) = \begin{bmatrix} \tilde{\rho}_{--} & \tilde{\rho}_{-+} \\ \tilde{\rho}_{+-} & \tilde{\rho}_{++} \end{bmatrix} \tag{37}$$

This matrix corresponds to the single-particle mixed states. As it was mentioned above, a single QD in an ensemble can be considered as an open quantum system subjected to losses due to the tunneling coupling. One can introduce a single-particle master equation describing the "losses" of the particular QD, arising due to the interaction with the rest of the lattice in the following way

$$\frac{\partial \hat{\rho}}{\partial t} = -\frac{i}{\hbar}\left[\hat{H}_{eff}, \hat{\rho}\right] - \frac{1}{2}\{\hat{\Gamma}, \hat{\rho}\} \tag{38}$$

where $\{...,...\}$ means the anticommutator of two operators. The effective Hamiltonian and the operator $\hat{\Gamma}$ are the spatial differential operators in 4-space given by

$$\hat{H}_{eff} = \frac{\hbar}{2}\begin{pmatrix} \hat{\Delta} & -\hat{\zeta}_{-\kappa}^* \\ -\hat{\zeta}_{-\kappa} & -\hat{\Delta} \end{pmatrix} \tag{39}$$

$$\hat{\Gamma} = \frac{\hbar}{2}\begin{pmatrix} \hat{\Gamma}_{--} & \Omega \\ \Delta\Omega & \hat{\Gamma}_{++} \end{pmatrix} \tag{40}$$

where $\zeta_{-\kappa} = [\Omega_R(\mathbf{r}) + \Omega_R(\mathbf{r}')]/2$, $\Delta\Omega = [\Omega_R(\mathbf{r}') - \Omega_R(\mathbf{r})]/2$, $\Delta = \omega_0 - \omega + 2(\xi_1 - \xi_2)$ and

$$\hat{\Delta} = \Delta - \frac{(\xi_2 - \xi_1)a^2}{2}(\nabla_\mathbf{r}^2 + \nabla_{\mathbf{r}'}^2) - i\frac{(\xi_2 + \xi_1)a^2}{2}\mathbf{k}(\nabla_\mathbf{r} - \nabla_{\mathbf{r}'}) \tag{41}$$

$$\Gamma_{--} = i\xi_2 a^2(\nabla_\mathbf{r}^2 - \nabla_{\mathbf{r}'}^2) + \xi_2 a^2\mathbf{k}(\nabla_\mathbf{r} + \nabla_{\mathbf{r}'}) \tag{42a}$$

$$\Gamma_{++} = i\xi_1 a^2(\nabla_\mathbf{r}^2 - \nabla_{\mathbf{r}'}^2) - \xi_1 a^2\mathbf{k}(\nabla_\mathbf{r} + \nabla_{\mathbf{r}'}) \tag{42b}$$

The formulated model opens the way for modeling of different types of quantum antennas. For example, to accommodate the solution for a closed loop scheme, one needs to consider a finite-length QD chain with periodic boundary conditions at the ends. The current per unit cell arising in the QD array can be decomposed into two parts as $\mathbf{j}(\mathbf{r}) = \mathbf{j}_{pol}(\mathbf{r}) + \mathbf{j}_{tun}(\mathbf{r})$, where $\mathbf{j}_{pol}(\mathbf{r})$ is the polarization current, and $\mathbf{j}_{tun}(\mathbf{r})$ is the current due to the tunneling of the charge carriers between the dots. These currents are related to the density matrix as shown by

Equation (38) with Relations (39)-(42). The current components $\mathbf{j}_{pol}(\mathbf{r})$ and $\mathbf{j}_{tun}(\mathbf{r})$ have different physical origin and contribute to different regions of the frequency spectrum. The implementation of Rabi waves in nanoantennas is possible both in the optical and infrared/THz ranges. The polarization current $\mathbf{j}_{pol}(\mathbf{r})$ is a prospective for the optical range, while the use of the tunneling current $\mathbf{j}_{tun}(\mathbf{r})$ leads us to the low-frequency nanoantenna. The current $\mathbf{j}_{pol}(\mathbf{r})$ arises from the polarization of a single QD due to the interlevel transitions dictated by the operating electric field. This is an amplitude modulated oscillation with the frequency of the quantum transition $\omega$. Its physical nature corresponds to the Mollow triplet in the theory of resonant fluorescence[2,3]. However, in contrast to the ordinary Mollow case, the satellites correspond to non-harmonic oscillations with nonuniformly broadened spectral lines.

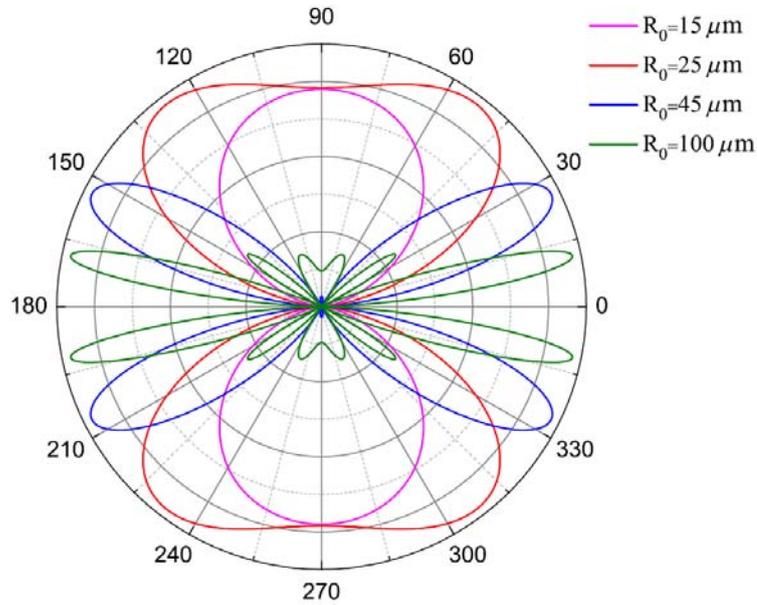

**Figure 7**. Radiation intensity pattern of a loop Rabi-wave nanoantenna. $\Omega_R = 10^{12}\,\text{Hz}$, $a = 12\,\text{nm}$, $k = -2\times 10^8\,\text{m}^{-1}$, $\Delta = -5\Omega_R$, $\xi_2 = \Omega_R$, $\xi_1 = 0.85\xi_2$ [126] Copyright (2013) SPIE.

The tunneling current corresponds to the intralevel drift of the particle along the chain. It comprises the partial contributions of the ground state ($\sim \xi_2$) and the excited state ($\sim \xi_1$) particles. Indeed, the spectrum of the tunneling AC current covers the vicinity of the Rabi frequency $\Omega_R$. For realistic QDs and background materials, this vicinity corresponds to the THz range. The tunneling current has no analogues in the standard theory of Rabi oscillations. The radiation pattern of a single loop Rabi-wave nanoantenna proposed in Reference[126] is similar to the radiation of an ideal magnetic dipole. The radiation pattern for a Rabi wave loop nanoantenna is shown in **Figure 7** for the different values of the loop radius[126]. One can see that the emitting characteristics of such antennas are electrically tunable via the variation of the intensity and phase velocity of the light wave.

The antenna emission in Reference[124] is considered as the interaction with a thermal photonic reservoir, which is described by the master equation

$$\dot{\rho} = \frac{i}{\hbar}[\hat{H}, \rho] - \frac{\Gamma(\omega)}{2}\left[\hat{\sigma}^+\hat{\sigma}^-\rho(t) - 2\hat{\sigma}^-\rho(t)\hat{\sigma}^+ + \rho(t)\hat{\sigma}^-\hat{\sigma}^+\right] \quad (43)$$

where

$$\Gamma(\omega) = \frac{d_{ab}^2\omega^3}{\pi\varepsilon_0\hbar} \int_0^\pi \sin\theta\cos^2\theta |F(\theta,\omega)|^2 d\theta; \quad F(\theta,\omega) = \frac{1}{l}\int_{-l/2}^{l/2} f(x) e^{-i\frac{\omega}{c}x\cos\theta} dx \quad (44)$$

The factor $\Gamma(\omega)$ plays the role of the parameter of dissipative coupling (see Section 2.2), however the value $F(\theta,\omega)$ looks like the radiation pattern of a classical antenna (with the wave function $f(x)$ of the antenna quasiparicle instead of the current density distribution[5]). It shows that the Lindblad term in Equation (43) describes the antenna emission.

For the spectrum of the radiation intensity, it is obtained

$$S(\omega; R, \theta) = \frac{k^4 d_{ab}^2 \sin^2\theta}{4\pi(4\pi\varepsilon_0)^2 R^2} \Gamma(\theta,\omega) \times$$

$$\left\{\frac{[\sin(\Psi)/\Psi]^2}{(\omega-\omega_0)^2 + (\Gamma/2)^2} + \frac{(3/4)[\sin(\Psi_+)/\Psi_+]^2}{(\omega-\omega_0-\Omega_R)^2 + (3\Gamma/4)^2} + \frac{(3/4)[\sin(\Psi_-)/\Psi_-]^2}{(\omega-\omega_0+\Omega_R)^2 + (3\Gamma/4)^2}\right\} \quad (45)$$

where $\Psi = kl(\cos\theta - \varphi)/2$, $\Psi_\pm = \Psi \pm (\Omega_R/\omega)\cos\theta$ and $\varphi$ is the phase shift per unit length.

The total radiation pattern given by Equation (45) is a sum of three elementary patterns $(\sin\Psi/\Psi)^2$ centered at the angles $\theta = \arccos(2\varphi/kl)$, $\theta_\pm = \arccos[2\varphi(1\pm\Omega_R/\omega)/kl]$. The weight of an elementary pattern is defined by the corresponding resonant line centered at one of the frequencies $\omega = \omega_0$, $\omega = \omega_0 + \Omega_R$, $\omega = \omega_0 - \Omega_R$, which corresponds to the Mollow triplet [2,3]. However, each triplet line corresponds to its own direction. The elementary radiation patterns for different values of the driven parameters are depicted in **Figure 8**. One can see that the pattern separation increases, and the patterns even attain complete separation (the angle of the maximum for the main lobe of one pattern coincides with the minimum for the other one). The value $\varphi = \varphi_{cr} = 1$ defines the regime of axial radiation for classical antennas [5]. For the phase shift $\varphi > 1$, the main lobe moves to the invisible region [5] and becomes unrelated to the observable angles $\theta$. In our case, there exists a specific critical shift for every partial diagram: $\varphi_{cr} = 1$ for the central line and $\varphi_{cr} = 1/\left(1\pm\frac{\Omega_R}{\omega}\right)$ for two side lines. These shifts depend on the field value. Putting it in a simple way, a single dipole antenna becomes as three "effective" antennas due to a strong coupling regime. Each "effective" antenna is characterized by its own emission frequency and radiation pattern controllable by the adiabatic field variation.

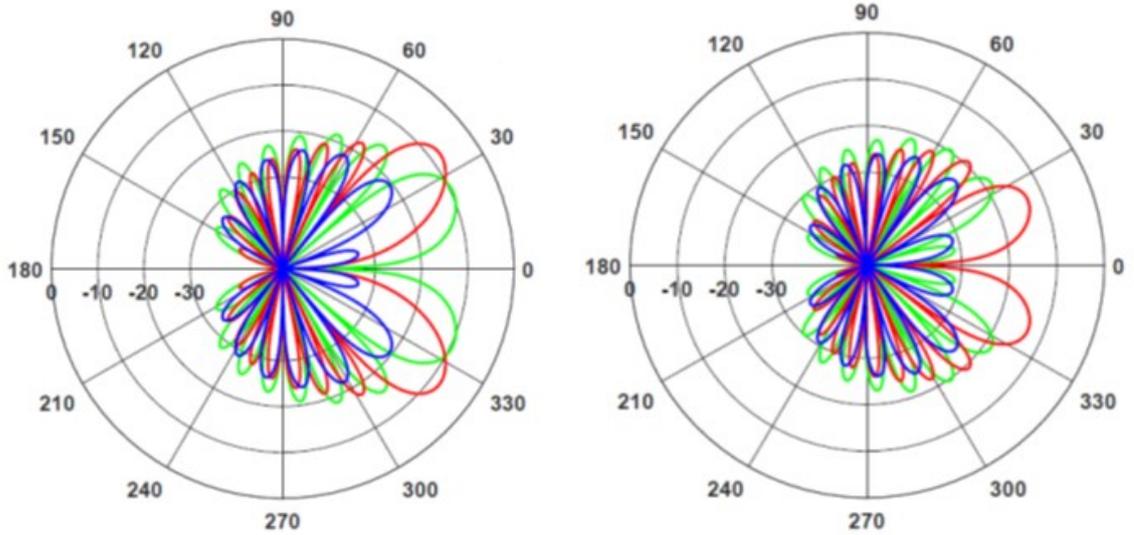

**Figure 8.** The elementary radiation patterns of the lines $\omega = \omega_0$ (colored red), $\omega = \omega_0 + \Omega_R$ (colored blue) and $\omega = \omega_0 - \Omega_R$ (colored green) on the logarithmic scale for different values of the phase shift. $kl/2 = 4\pi$, $\Omega_R = 0.2\omega$. (a) $\varphi = 1.0$. (b) $\varphi = 1.2$ [124].

Some other tools for the control of the antenna radiative properties may be presented by using DC electric field. Quite a long time ago, Bloch and Zener described the phenomenon of Bloch oscillations of a single particle in a periodic potential under the influence of a static force [127,128]. As it was predicted in References [32,129,130], the simultaneous Bloch oscillations and resonant interaction with the AC component lead to the appearance of a combination of Rabi and Bloch oscillations, strongly influencing one another (so-called "Rabi-Bloch oscillations" [32,129,130]). This behavior manifests itself in multiline spectra with different combinations of Rabi and Bloch frequencies, which opens the new ways for electrical control via adiabatic variance of AC field and DC voltage.

### 3.3. Non-linear antennas based on Rabi-solitons.

A significant role for building the non-linear antennas belongs to the dipole-dipole (d-d) electron-hole interactions inside the isolated QD [125,131–141]. Such interactions produce some additional (depolarization) field with respect to the incident one. It makes the total field consistent with the particle motion (the charge carrier moves in the field induced by themselves, but not in the incident one). In particular, the d-d interactions induce a fine structure of the QD absorption-emission spectrum with the splitting $\Delta\omega$ of the spectral lines, calculated in Reference [134] for disk-like and cigar-like spheroidal QDs. The sign of splitting is opposite for the absorption and emission, while the splitting value is the same. Some of the d-d effects, such as bifurcation and essentially anharmonic Rabi oscillations regimes, were predicted both for classical and quantum light [131–133]. Thereafter, they have been experimentally observed with a good consistency between the experimental data and the theoretical predictions [136–138]. The phenomenological description of this interaction is based on the concept of local fields [125].

Obviously, in quantum electrodynamics, the d-d interaction exists due to the exchange between the electrons and the holes by the virtual vacuum photons [134,139,140]. A simplified analysis of the d-d effects can be performed using the phenomenological Hartree–Fock–Bogoliubov approximation [133]. A more regular approach was developed by means of BBGKY (Bogoliubov–Born–Green–Kirkwood–Yvone) –hierarchy [139]. The main quantum optical manifestation of the d-d interaction is the so-called photonic state dispersion in excitonic metamaterials [139]. In particular, it is manifested as different refractive indexes for the single photons and coherent states of light [139].

The nanoantennas can be realized with the self-trapping discrete solitons on the nonlinear metasurfaces built of self-organized semiconductor QDs coupled simultaneously due to the interdot tunneling and the d-d interaction [125,141]. In this context, dynamically stable soliton structures (fundamental discrete solitons, vortex solitons, and breathers) provide attractive mechanisms for the design of the controllable antennas with various geometric shapes [125,141]. To take into account the d-d interaction, the component $\Delta \hat{H} = \hbar \sum_j \left( \langle \hat{\sigma}_j^- \rangle \hat{\sigma}_j^+ + \langle \hat{\sigma}_j^+ \rangle \hat{\sigma}_j^- \right)$ should be added to the Hamiltonian (36). As a result, in References [125,141], the system of 2D second-order discrete non-linear Schrodinger equations with the cubic non-linearity was obtained for the probability amplitudes $A_{p,n}(t), B_{p,n}(t)$ of the ground and excited states, respectively ($p,q$ is the set of QD numbers in the lattice). Using numerical techniques, the localized solutions of the corresponding equations have been constructed; their stability and mobility have been tested. The Rabi solitons induce the spatially distributed electric polarization, which is proportional to $S_{p,q} = A_{p,q} B_{p,q}^*$ for QDs with the numbers $p,q$.

The far-field structure produced by the near field is characterized by the radiation pattern

$$F(\theta,\varphi;\phi_1,\phi_2) =$$
$$= \sin\theta \sum_{p=-N_1/2}^{N_1/2} \sum_{q=-N_2/2}^{N_2/2} S_{p,q} e^{ip(\frac{\omega}{c} a \sin\theta \cos\varphi + 2\phi_1)} e^{iq(\frac{\omega}{c} a \sin\theta \sin\varphi + 2\phi_2)} \quad (46)$$

where $\theta, \varphi$ are the meridional and azimuthal angles, $\phi_{1,2}$ are the phase shifts over the period in the orthogonal directions, $N_{1,2}$ are the numbers of QDs in the orthogonal directions. The 2D electromagnetic radiation patterns induced by the given polarization profiles suggest considering the present system as transmitting antennas. The driving field plays the role of an external energy source. For the data transfer, the emitted field should be temporally modulated by the input signal through the driving electromagnetic field. Actually, this modulation is relatively slow, and, therefore, can be considered as an adiabatic process.

The phase variation allows one to scanthe space by turning the main lobe of the radiation pattern. One of the most efficient means to implement the antenna control is the use of the switched-beam systems able to impose the different angular patterns in order to enhance the emitted power in a preferable direction. The concept of the solitonic nanoantenna, presented in References [125,141], is based on the beam-forming algorithm, which is determined by the set of factors $S_{p,q}$, using the soliton-emission profile (see Equation (46)). As several different stable solitons may exist in the given array, the choice of the overall profile depends on the initial conditions. This implies that a finite number of predefined radiation patterns exists for a given set of the antenna parameters, while the initial conditions give an opportunity for choosing and switching the suitable one. The partial radiation patterns in the *E*- and *H*-planes for different types of the solitons are presented in References [125,141].

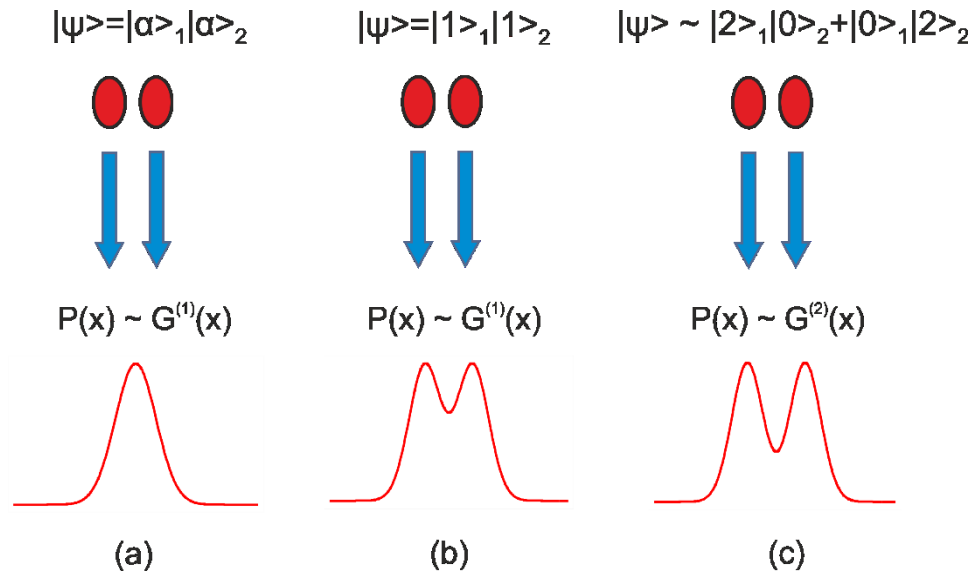

**Figure 9**. Near-field imaging of the two-mode source in different quantum states: a) coherent state $|\alpha>_1 |\alpha>_2$, with the unit amplitudes, $\alpha = 1$; b) single-photon Fock states $|1>_1 |1>_2$; c) NOON state $|\varphi> \propto |2>_1 |0>_2 + |0>_1 |2>_2$.

## 4. Applications of Quantum Antennas

As it has already been discussed here, the quantum antennas can produce non-classical states of light and shape both the temporal and spatial structure of the field in the way impossible with classical emitters. This implies possible applications of the quantum antennas in super-resolving imaging and far-field sensing (for example, as radars).

### 4.1. Quantumness of the emitted field: manifestations

Before discussing the quantum antenna applications, it is useful to show how the quantumness of the field emitted by an antenna manifests itself in conventional measurements, for example, in the measurements of the field correlation functions of different orders.

Let us consider the simplest measurement set-up for resolving two slit sources (see **Figure 9**). For simplicity's sake, let us assume that each source is monochromatic and produces a single linearly polarized spatial mode, with the source frequencies and the polarizations being identical. In this way, the field amplitude operator at the point $x$ of the imaging plane (on the line perpendicular to the slits) can be represented as

$$E(x) = PSF(x - d)\hat{a}_1 + PSF(x + d)\hat{a}_2, \qquad (47)$$

where $\hat{a}_{1,2}$ are the photon annihilation operators of the first and the second sources, placed at positions $d$ and $-d$, correspondingly, in the source plane; the source plane is assumed to be parallel to the image plane. The function $PSF(x)$ is the point-spread function describing the field propagation from the thin slit source to the image plane. For our example, let us take the simplest *sinc* function (typical for a thin slit source [142]), $PSF(x) \propto sin(kx)/kx$, where $k$ is the multiplier, proportional to the module of the wave vector of the emitted field and dependent on the set-up parameters.

If the states of our sources are uncorrelated and classical (coherent), according to Equation (47), the measurement of the intensity in the image plane gives

$$G^{(1)}(x) \propto |PSF(x - d) + PSF(x + d)|^2.$$

One can see in the left panel of **Figure 9**, that this measurement cannot lead to visual resolution of the slits spaced $2d=1.5/k$ apart (this distance is also assumed for the other measurements of our example). Now, let us assume that the sources emit single-photon states in the respective modes. Then, from Equation (47), one gets

$$G^{(1)}(x) \propto |PSF(x - d)|^2 + |PSF(x + d)|^2,$$

and the sources can be resolved using their image (see the central panel in **Figure 9**). One can further increase the resolution by measuring the higher-order correlation functions for the other states. For example, in the right panel of **Figure 10,** one can see the results of the $G^{(2)}(x,x)$ measurements for the two-photon NOON state, $|\varphi> \propto |2>_1 |0>_2 + |0>_1 |2>_2$.

Thus, devising the measurement and the source state, one can indeed increase the resolution of the given object (which is actively researched now in the intensively developing field of quantum imaging [142–144]).

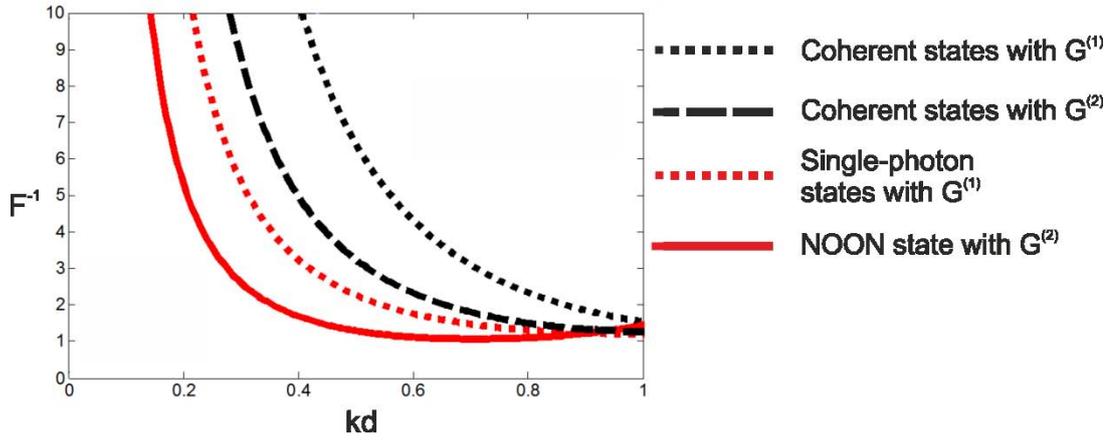

**Figure 10.** Inverse Fisher information for the estimation of the distance, *kd*, between the slits for the different states of the two-slit source and the different measurements.

Notice that rather an empirical concept of the "object resolution" can be formalized and quantified by introducing the concept of informational content of an image. If the object is described by the set of parameters $\vec{x}$, and the image is described by the complete set of the probabilities $p_j(\vec{x})$, satisfying $\sum_{\forall j} p_j(\vec{x}) = 1$, the amount of information in the image about the parameters $\vec{x}$ can be described by the Fisher information matrix [145]

$$F_{mn}(\vec{x}) = \sum_{\forall j} \frac{1}{p_j} \left[\frac{\partial}{\partial x_m} p_j\right]\left[\frac{\partial}{\partial x_m} p_j\right].$$

(48)

The practical value of the Fisher information matrix is in its usefulness for providing minimal attainable errors of an unbiased estimate of the parameters via Cramér–Rao bound [146,147]

$$\Delta_m^2 \geq \frac{1}{N}[F^{-1}]_{mm},$$

(49)

where $\Delta_m^2$ is the variance of the estimate of the parameter $x_m$, and $N$ is the number of the measurement runs. Notice that there are estimation methods allowing one to attain the equality in the bound (49) (for example, the maximal likelihood estimation can do it [148]). As it is shown in **Figure 10**, for our example, the increase in resolution shown in **Figure 9**, directly corresponds to the increase of the Fisher information about the slit separation.

The "classical" Fisher information (48) depends both on the measurement set-up and on the state of the imaging source. To separate these influences, one assumes that the probabilities are generated by the set of measurements described by the positive-operator valued measure (POVM) via the Born's rule:

$$p_j = \text{Tr}\{\hat{\Pi}_j \hat{\rho}\},$$

(50)

where $\prod_j$ are the operator elements of the POVM and $\hat{\rho}$ is the density matrix of the radiation after being transformed by the object. Then, let us seek to maximize the *m*-th diagonal element of the Fisher information matrix (48) over all the possible measurements,

$$\Phi_{mm} = \max_{\forall \text{POVM}} [F_{mm}(\vec{x})] . \tag{51}$$

It is possible to show that this maximum is attained for the so-called "quantum Fisher information" matrix [149], defined as

$$\Phi_{mn} = \frac{1}{2} \text{Tr}[(\hat{L}_m \hat{L}_n + \hat{L}_n \hat{L}_m) \hat{\rho}], \tag{52}$$

where the operator for the symmetric logarithmic derivative for the *m*-th parameter is defined in the following way $\partial_m \hat{\rho} = \frac{1}{2}(\hat{L}_m \hat{\rho} + \hat{\rho} \hat{L}_m)$. The quantum Fisher information bounds from above the information about the parameters attainable for a given imaging state. And for quantum states, this information can vastly exceed the information attainable with the help of classical or non-quantum correlated states [150]. Using entangled states, one can even attain the ultimate limit on precision, the so-called "Heisenberg limit" (which is the "holy Grail" of quantum metrology [151]). However, here, one must add that it is not always possible to devise a measurement able to realize such an informational advantage.

The quantumness of the states emitted by quantum antennas can be manifested in a more subtle way than those described above. With quantum states, one can get the measurement results impossible for classical states. One of the most spectacular demonstrations is the violation of the so-called Bell (or the so-called Bell-type) inequalities for the probabilities of the measurements on different parts of the correlated quantum systems [152,153]. Let us demonstrate how the classically impossible measurement results arise for our example with a two-mode source, producing a two-photon NOON state as it was described in Reference [154].

We coherently shift both modes and direct them to the simplest bucket detectors able to distinguish only between the presence and the absence of photons. For simplicity's sake, we assume them to be ideally efficient. In that way, our measurement realizes the projections of the NOON state on the pair of coherent states. The probability $P_j(x)$ to register a "click" for the *j*-th mode, coherently shifted with the amplitude $x$, is

$$P_j(x) = 1 - \frac{1}{2} e^{-|x|^2} \left( \frac{1}{2} |x|^4 + 1 \right), \tag{53}$$

and the probability to have "clicks" on both detectors for the first mode shifted by $x$ and the second mode shifted by $y$ is

$$P_{12}(x, y) = 1 - P_1(x) - P_2(y) + \frac{1}{4} e^{-|x|^2 - |y|^2} |x^2 - y^2|^2 . \tag{54}$$

For the measurements with two different shifts, $x_{12}$ and $y_{12}$, for the corresponding modes, one can construct the so-called Clauser-Horne inequality [155] from the probabilities (53,54):

$$-1 \leq P_{12}(x_1, y_1) + P_{12}(x_1, y_2) + P_{12}(x_2, y_1) + P_{12}(x_2, y_2) - P_1(x_2) - P_2(y_1) \leq 0. \tag{55}$$

The relation would always hold for classically correlated states of the modes 1 and 2. For our 2-photon NOON states, the inequality (55) can be violated: the central part can be less than $-1$ [154].

Notice that by measuring the correlation functions for different distances from the source, one can even determine the degree of non-classicality, in particular, the entanglement [156].

**4.2. Quantum antennas for field shaping**

In Section 2, we have already remarked the possibility of shaping intensity and higher-order correlations of the field, emitted by the antenna, by designing the initial states, geometry and driving of the antenna [66,70,83,84]. In particular, the correlations of the emitted field can be widely varied by changing the initial state of the antenna, and the desired spatial distribution of the correlations can be achieved by designing the antenna state [66]. Creating the initial timed Dicke states of the antenna, one is able to achieve highly directional emission of a single photon [80]. Curiously, a large number of emitters in the antenna by itself does not guarantee high directivity of the given correlation function. For example, a few tens of emitters for a two-photon Dicke state of a non-interacting linear array quantum antenna is already sufficient for very high (few degrees) directivity of $G^{(2)}$ [66]. However, for the same array antenna of an interacting TLS, one needs hundreds of emitters to reach high-intensity directivity for a completely excited initial state [157].

Notice that multi-particle entangled states are quite difficult to produce. Creating even few-excitation Dicke states of an antenna is challenging. However, there is a way of emulating the effects of a Dicke entangled state even for a linear array of non-interacting incoherent sources (even thermal ones [94]).

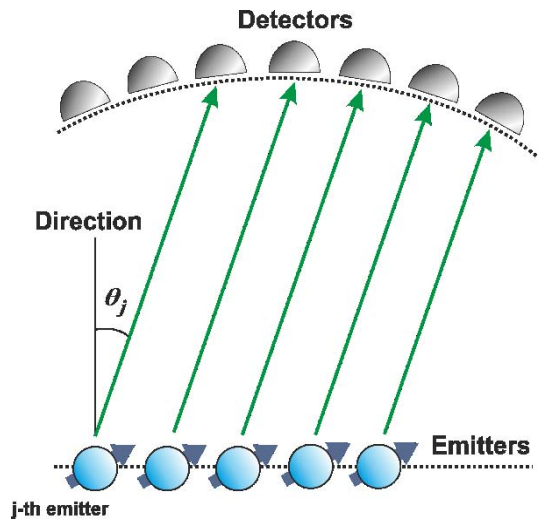

**Figure 11.** The set-up for measuring the *m*-order spatial intensity correlation function for the field emitted by the array antenna considered in Reference[94].

Let us consider the set-up depicted in **Figure 11**. The field, emitted by the $N$-emitters linear array antenna in the far-field zone, is registered by $m$ detectors placed at the same distance from the antenna. For the initially completely excited state of all $N$ emitters, the probability to register a photon at the direction $\theta_2$, when $m$-$1$ photons are registered at the direction $\theta_1$, is proportional to the $m$-order correlation function

$$G^{(m)}(\theta_1, \ldots, \theta_1, \theta_2) \propto \frac{N-m}{N} + \frac{m-1}{N^2}\left[\frac{\sin N\delta\varphi_{12}}{\sin \delta\varphi_{12}}\right]^2, \tag{56}$$

where $\delta\varphi_{12} = 0.5kd(\sin\theta_2 - \sin\theta_1)$ is the angle, $d$ is the distance between the TLS emitters, and $k$ is the wavenumber. For the photon emitted in the direction $\theta_2$, Equation (56) shows the same dependence on the angle $\theta_2$ as for the $N$-emitter antenna initially in the Dicke state with $N$-$m$+$1$ excitations [158]. The same takes place even for the TLS being replaced with statistically independent incoherent thermal light sources. Then, one has the following expression instead of Equation (56)

$$G^{(m)}(\theta_1, \ldots, \theta_1, \theta_2) \propto 1 + \frac{m-1}{N^2}\left[\frac{\sin N\delta\varphi_{12}}{\sin \delta\varphi_{12}}\right]^2. \tag{57}$$

The reason for this seemingly paradoxical phenomenon is the intrinsic quantum nature of the photon detection for the case. The measurement of $m$-$1$ photon at some angle actually projects the antennas state on the Dicke state with $N$-$m$+$1$ excitations. The result described by Expression (57) was actually confirmed by the experiment [94].

Thus, shaping of the emitted field can be accomplished in a "post-processing" way, by designing the quantum measurements able to project the initial state on the state providing the necessary directivity.

### 4.3. Generation of non-classical states

The quantum antenna as a set of a finite number of quantum emitters is a natural producer of non-classical states. A single completely excited TLS emits just one photon, and if it is emitted in a single-mode field, one has a single-mode single-photon Fock state. Applying the same simple logic for the set of $N$ TLS, one arrives at a conclusion about the possibility of $N$-photon Fock state generation. It is extremely important resource: for example, Fock states allow one to reach optical channel capacity in quantum communications [159], or the Heisenberg limit in interferometric phase measurement [150].

However, the spontaneous emission occurs spontaneously. Differently placed emitters emit photons in different moments of time, and, generally, in different directions. Even for a single TLS, the different emitted photons from consecutive cycles of excitation-emission would be "spread" over the continuum of reservoir field modes in a different way. Generally, these photons will be distinguishable. Even for just single-photon generation, the long researches were trying to approach reliable generation of the single-photons suitable for quantum information processing and communications due to non-ideality of feeding emitters, the emitter itself collecting and emitted field collecting systems [160]. Only very recently, the top-end quantum dot-based sources have succeeded in demonstrating simultaneously the purity exceeding 99% and indistinguishability of single photons exceeding 95% [161].

As it has already been demonstrated here, quantum antennas are able to shape the emitted radiation, provide quite high directivity just by the interference effects, and even provide the direction-dependent photon-number statistics[84,85,94,158]. It was shown that one can generate few-photon states with the given pattern of spatial correlations by designing the initial state of the antenna (e.g. co- or contra-correlated bi-photons)[66]. The antenna can potentially be a very good source even for single photons[80]. Of course, it is quite challenging to create multi-particle entangled states. However, in this case, one can use the very effect that hinders multi-photon excitation of the antenna: the dipole blockade discussed in Subsection 2.2[68].

Also, the same feature, making the emitted photons distinguishable, can be exploited to create the entanglement[162]. Indeed, let us have a look at the single-photon state (see Expression (12)) created as a result of the TLS spontaneous decay in the vacuum of the reservoir. For large times, when the TLS is in the ground state, the emitted field decouples from the TLS and can be described by the wave function $|\varphi> \propto \sum_{\forall k} c_k(t)|1>_k$. Here, the index $k$ numbers all the wave vectors and polarizations. Let us divide all the modes in groups 1 and 2 in such a way that $\sum_{\forall k\,1}|c_{k1}(t)|^2 = \sum_{\forall k\,2}|c_{k2}(t)|^2 = |C|^2$ and introduce a single-photon wave-packets

$$|1_j> \propto \frac{1}{|C|}\sum_{\forall k\,j} c_{kj}(t)|1>_{kj}, \qquad (58)$$

where the vectors $|1>_{kj}$ describe the state with the photon in $k$-mode of the $j$-th group. Then, the total field state can be represented as a single-photon NOON state

$$|\varphi> \propto |1_1>|0_2> + |0_1>|1_2>, \qquad (59)$$

where the vectors $|0_{2j}>$ describe the vacuum state of the $j$-th group of the modes. The state (59) is the maximally entangled one, and its non-classicality can be demonstrated with the measurement akin to the one described in Subsection 4.1 with the violation of the inequality (55).

A single photon can indeed be entangled with itself. And this effect can be implemented to produce streams of such "self-entangled" photons by collecting the emitted photons with two nanoantennas, even when exciting a single quantum emitter with an incoherent pump[162]. Notice that one can still get some non-classicality for multi-photon emitted light by simply exciting a group of independent emitters despite the irregularity of their placing and randomness of emissions[163,164]. Finally, it is useful to note that, as it was also discussed in Subsection 2.6, the interactions between the emitters of the antenna are also able to produce non-classical states of emitted light even for initially uncorrelated states of the emitters or classical driving[24,64,71,101,108,165]. However, designing "talking antennas" for emitting particular states is rather a challenging task involving multi-parametric nonlinear optimization[66].

**4.4. Quantum imaging**

Naturally, possible implementations of quantum antennas for imaging and sensing are connected with their ability to emit non-classical and non-classically correlated multi-mode field. The underlying logic is illustrated in Subsection 4.1. However, one must mention here that the quantum-information approach also brought quite unexpected and spectacular advances in imaging even for classical antenna devices, namely, the localized incoherent sources. It became possible to see an age-old problem of resolving two bright points in a far-field zone in a different light.

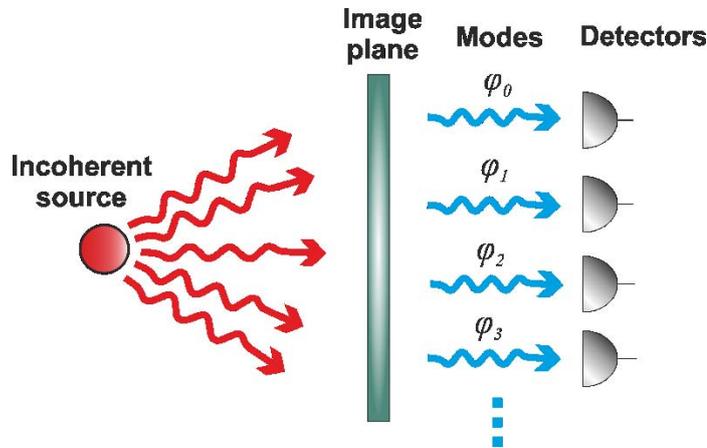

**Figure 12**. The set-up for "dispelling" the "Rayleigh's curse" for incoherent imaging. Here, $\varphi_l$ is the Hermite-Gaussian mode of the *l*-th order [172].

Indeed, it is school-physics knowledge that it may not be possible to resolve two close point sources(see **Figure 9**). Ultimately, the value of the wavelength defines the resolution limit for the case [166]. And this is directly translated into the informational content of the measurement (see **Figure 10**). One can see that for a smaller distance between the slits, the inverse Fisher information grows and, eventually, the lower bound for the error (48) diverges as $(kd)^{-1}$. This phenomenon has recently been aptly named the "Rayleigh's curse" or "Rayleigh's catastrophe" and discussed in details [167–175]. It was understood that the "curse" is not universal. It actually stems from the particular form of the measurement performed on the field arriving from the source. One can dispel the "curse", for example, by using multi-photon quantum correlated states and the so-called "centroid measurements" [168]. But even for incoherent independent sources, the quantum Fisher information appears to be constant and independent on the separation between the sources [169,171]. This information can be accessed, for example, by decomposing the field in the imaging plane in terms of Hermite-Gaussian modes (performing spatial-mode demultiplexing) and measuring them separately (see **Figure 12**) [172].

However, the two-emitter antenna is rather a special source for emitting tasks. One can actually "dispel" the "curse" only for the estimation of a single parameter, such as the distance between the sources,. The "curse" is generally back [176,177] for two and more parameter estimation (for example, distances between different emitters of the antenna),. The Fisher information matrix (48) inevitably becomes singular. One cannot say anything about the point apart from the fact that it is a point (i.e. that the size of the object is zero).

Let us demonstrate the "curse" and its absence by detecting the rotation angle of the two-emitter "talking" antenna considered in Subsection 2.2. We consider the scheme depicted in **Figure 3**. The second-order correlation function $G^{(2)}(\theta_1, \theta_2; t, t+\tau)$ of the emitted field in the far-field zone is proportional to the probability of detecting the first photon in a small solid angle at the direction $\theta_1$, and the time *t* (actually, in a small time-interval near *t*), and the second photon at the direction $\theta_2$ and the time $t+\tau$. For simplicity's sake, let us assume that we catch the photons flying in the same direction, $\theta$, in the plane perpendicular to the dipole moments of the antenna emitters. So, from Equation (25), one has the following result for the probability to detect both photons from the antenna rotated with the angle $\delta\alpha$

$$p(\theta,\delta\alpha,t,\tau) \propto exp\{-\gamma_{11}\tau\}[(1 + [\cos\varphi(\theta + \delta\alpha)]^2)\cosh\gamma_{12}\tau - 2\sin\varphi(\theta + \delta\alpha)\sinh\gamma_{12}\tau + [\sin\varphi(\theta + \delta\alpha)]^2 \cos f_{12}\tau]. \qquad (60)$$

If we perform just one measurement, with the detectors being far from the antenna, the probability (60) is small, $p(\theta,\delta\alpha,t,\tau) \ll 1$. Then, the Fisher information (48) for the case is

$$F \approx \frac{1}{p(\theta,\delta\alpha,t,\tau)}\left[\frac{\partial}{\partial\delta\alpha}p(\theta,\delta\alpha,t,\tau)\right]^2. \qquad (61)$$

Immediately, one sees that the probability to catch both photons for zero delay, $\tau = 0$, does not depend on the observation and the rotation angles, and the Fisher information is zero. The measurement is not informative despite the fact that the probability of registering two photons is not zero. However, for a non-zero delay, the Fisher information (61) is not zero even for $\delta\alpha = 0$. Moreover, the information survives even for a non-sharp detection, when the time window of the second detector, opened after registering the first photon, is rather large[165]. Notice that the "dispelling" of the "curse" for the traditional example of two slits, illuminated by the two-emitter quantum antenna, can also be done by the measurement with a finite delay[165].

### 4.5. Quantum radar

Radar is a far-field sensing set-up that uses antennas to irradiate an object and detects the field scattered by the object to infer the information about it. Radars are used for object (target) detection, i.e. for the determination of whether the target is present or absent inside a given solid angle. Also, radars are used to determine the distance to the target and its velocity, and, ultimately, for the target configuration recovery[178].

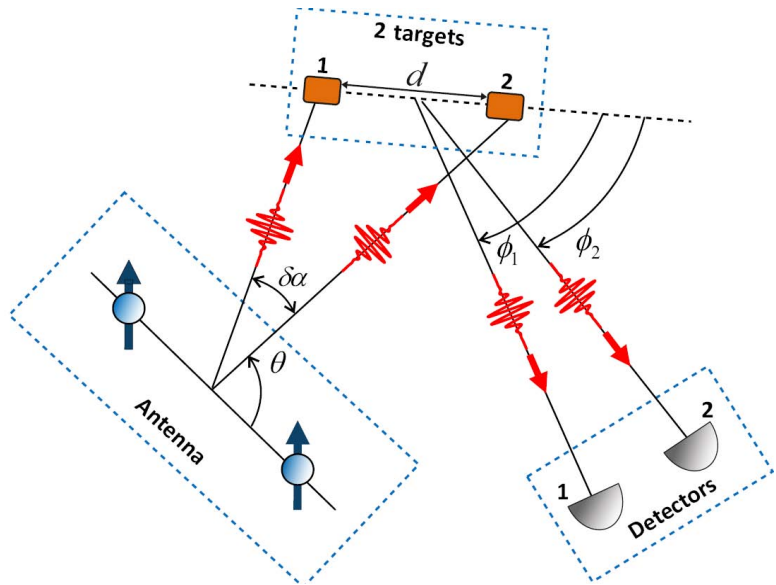

**Figure 13**. The "quantum noise radar" scheme for inferring the distance, *d*, between two small targets, 1 and 2.

Quantum correlations of the field produced by the antenna can be used for enhancing the performance of radars. The idea of using entangled states to illuminate the target gave birth to a subfield of research named "quantum radars" [179–182]. By retaining a part of the generated entangled state and sending the other part toward the target, and correlating the scattered field with the retained part, one can increase the resolution of the radar, in particular, its ability to discern the scattered signal in a noisy reservoir. Such an advantage of the so-called "quantum illumination scheme" [183–185] was experimentally confirmed [186,187]. Similar advantages can be obtained with the "ghost imaging", when the scattered field is registered with the simplest binary detector (i.e. the one able only to distinguish between the presence and the absence of the signal), with the retained part being scanned [188]. The quantum illumination strategy was also shown to be feasible in the microwave region, where the losses and noise are lower in comparison with the near-optical region [189]. Curiously, the "ghost imaging" can be also performed with classically correlated fields, but with a lower signal-to-noise ratio [190]. Such a "quantum inspired" imaging scheme resembles the well-known classical "noise radar", where the target is illuminated by noise, and the delayed copy of this signal is used to be mixed with the scattered field to study the correlation of both noises [191]. For the delay corresponding to the distance to the target, the intensity correlation for these noises would exhibit a sharp maximum. It is interesting that such a classical scheme can be enhanced by using a two-mode squeezed field for target illumination when the correlations of the measured intensities are studied [192].

To illustrate the resolution advantages achievable by using quantum antennas, let us consider a recently developed far-field sensing scheme for the target configuration recovery [165]. The scheme is shown in **Figure 13**. The field, generated by the two-TLS antenna considered in subsections 2.2 and 4.3, illuminates two small scatterers, and the scattered photons are registered on the detectors placed in the directions $\phi_1$ and $\phi_2$ from the target. If we catch both photons at the same direction, $\phi = \pi/2$, , one would get exactly the result (60) for the probability to detect the photons at the moments *t* and *t+τ* with the angle $\delta\alpha \approx 2d/R$ (R is the distance from the antenna to the target, and d is the distance between the scatterers). Thus, the "Rayleigh's curse" is successfully lifted, and the super-resolution for $d \to 0$ can be reached [165]. Notice that for zero delay, *τ=0,* the Fisher information (61) is zero for $d = 0$ even for different observation directions $\phi_1$ and $\phi_2$.

It is interesting that very recently the measurement of the delayed intensity correlation function for the bi-photon entangled illumination field has been shown (both theoretically and experimentally) to provide large gain in comparison with uncorrelated illumination field for the target detection [193].

## 5. IMPLEMENTATION AND MATERIALS FOR QUANTUM ANTENNAS

In this section, we will give a brief outlook of some real physical structures and systems corresponding to the theoretical models considered above. Of course, it is hardly possible to provide a comprehensive review of all the systems potentially able to act as quantum antennas In fact, practically anything able to be excited and emit photons can be a possible candidate. First of all, we concentrate our attention on the physical systems allowing for separation, manipulation and selective excitation of quantum emitters, i.e. on the systems that can be arranged in an antenna array to exploit the interplay of quantum and classical correlations. To name a few, these are plasmonic metals (such as gold, silver, aluminum, copper[194]), graphene[195], carbon nanotubes (CNTs)[196-202], self-assembled quantum dots (QDs) (fabricated by molecular beam epitaxy on a semiconductor substrate[203]), color defects in diamonds[204-2010], cold atoms and ions in traps[211,212], arrays of superconducting qubits[213], defects in 2D semiconductors[214], and organic molecules[215,216]. With the current state of nanotechnologies, these objects can be produced/placed with nanometer precision.

So, mostly for illustrative purposes, here we just choose several practical and realizable examples of solid-state emitters allowing for building quantum antennas and for manifestation of the specifically quantum effects discussed in this review, namely, non-classicality and quantum correlations of emitted radiation, collective emission effects, and quantum correlations between emitters. Namely, these are plasmonic nanoantennas, graphene structures, quantum dots and color centers in dielectrics.

Concerning receiving antennas, all the referred systems may be considered as potential candidates also for photon detection. However, the emission and absorption properties of the quantum emitters can be rather different, and the reciprocity theorem breaks down in many cases[35,101]. As a result, in contrast with the classical antennas, the properties of the given configuration as a transmission and receiving antenna are not identical and should be studied separately.

Also, we discuss such practical aspects of the quantum-antenna parameters as frequency-dependence and antenna matching with the emitter/load. Then, we briefly discuss the problems of electromagnetic compatibility (EMC) at the nanoscale.

### 5.1. Nanoantenna configurations in microwave, THz, infrared optics

The most simple antenna geometry is the system of two end-to-end aligned rods separated by a small gap (see **Figure 14a**). The gap corresponds to the area of dramatic field enhancement, therefore, it is the optimal one to place a detector (or a generator). A single wire can be viewed as a dipole fundamental building block of more complex antennas, in particular, Yagi–Uda antennas (see **Figure 14b**)[217–220]. More complicated antenna geometries were studied over the last years. Among them are the bow-tie antennas[221,222] (**Figure 14c**) and the cross antennas[43,223] (**Figure 14d**). One of the recent tendencies is the study of more complicated antennas, such as combined particles of semiconductor quantum dots on the dielectric background (**Figure 14e,f**) or a semiconductor QD inside a quantum wire of another material, etc.[224–226].

The problems of implementation of the optical antenna result from a small size and the resonant properties of nanostructures. The characteristic dimensions of such an antenna are of the order of the radiation wavelength, and for optical antennas this requires fabrication accuracies down to a few nanometers[117]. As a consequence, electrons in nanostructures of different types do not respond to the wavelength $\lambda$ of the external source but to an effective wavelength $\lambda_{eff} \ll \lambda$, which is defined by the dispersive properties of the charge carriers. For example, in conductive CNTs $\lambda_{eff} \cong \lambda v_F / c$, where $v_F = 10^{-2} c$ is the Fermi velocity of $\pi$–electrons [196–202]. The strongly reduced wavelength of the modes of plasmonic wires makes them suitable for the applications in different spectral regimes. The spatial extent of a detector or a source is commonly much smaller than the wavelength of the radiation, and is typically of the order of $\lambda/100$ to be ~5 nm for the optical regime [117]. For this length scale, the light–matter interaction becomes quantized, and the sources and receivers become quantum objects, as it was mentioned above.

Physics of nanoantennas has its roots in nanoplasmonics. For an increasing frequency of the excitation, the conductive electrons exhibit an increasing oscillation amplitude as well as an increasing phase lag. As soon as the phase lag approaches $\pi/2$, the amplitude of the charge oscillation goes through the maximum and is only limited by the internal (Ohmic and radiation) damping of the system. In metallic nanoparticles, this resonance corresponds to the localized plasmon resonance, which, for certain materials (such as gold, silver, aluminum, copper, graphene, CNT), happens to appear within or close to the visible spectral range. Plasmon resonances may be exploited to impact the drawbacks of traditional antenna systems in this frequency range (such as enhanced Ohmic losses compared with the radio-frequency regime). The major challenge in the design of macroscopic antennas is the impedance matching between the antenna and the external source, which is considered in detail in Reference [5]. The concept of the antenna impedance is applied to optical antennas [117]. The value of the antenna impedance is directly expressed through the photonic density of states [117].

The great progress in nanotechnologies has opened up different ways for the implementation of the antennas transmitting and receiving light signals at the nanoscale. This length scales became increasingly accessible as the novel tools, such as focused ion beam milling [227], electron-beam lithography [228] and nano-imprint lithography [229] have been improved.

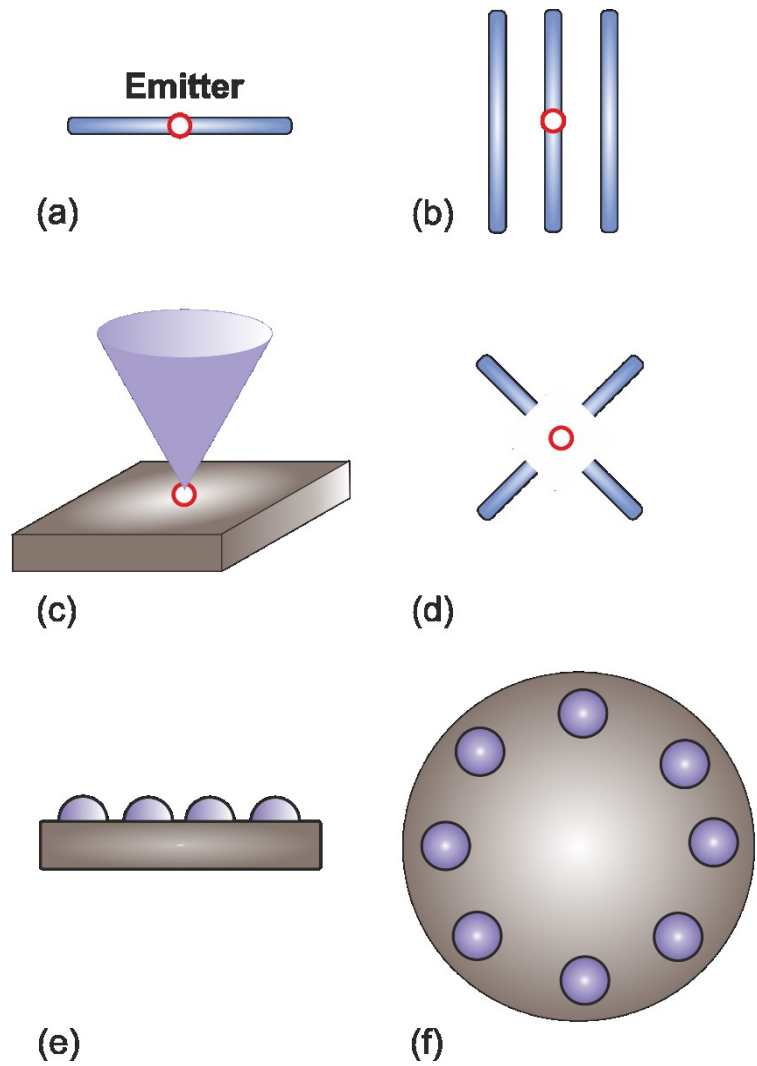

**Figure 14.** Examples of the simplest antennas configurations: a) wire dipole antenna corresponding to the aligned rods separated by a small gap, with the emitter placed inside it (denoted by the red circle); b) Yagi-Udo antenna (the system of dipoles, one of which is active and all the others are passive) [217–220]; c) bow-tie antenna (conductive cone perpendicular to the conductive background); d) crossed dipole antenna; e) 1-D array of semi-spherical semiconductor QDs placed at the plane dielectric background; f) the periodic loop of QDs placed at the plane dielectric background (top view).

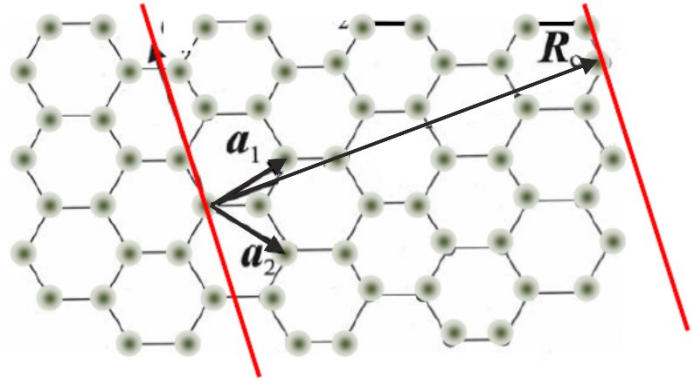

**Figure 15.** Crystal structure of graphene and its rolling to CNT; $\mathbf{a}_{1,2}$ are the unit vectors of the crystal lattice, $\mathbf{R}_c$ is the radius vector between the identical points after rolling.

## 5.2. Carbon-based materials

Speaking about carbon-based media, we will mean mainly two materials: graphene and CNTs. Their general potential for the antenna applications is defined by the corresponding unique electrical and optical properties, which are determined by their configuration [196–202,230–232]. A single-wall CNT-structure may be presented as a graphene sheet rolled to a cylinder over the red parallel lines as it is shown in **Figure 15**. The vector may be presented as $\mathbf{R}_c = m\mathbf{a}_1 + n\mathbf{a}_2$, where (*m,n*) are some integers, which completely characterize the CNT configuration. The CNTs (*m,m*) are named "armchair" tubes, the CNTs (*m,0*) are named "zigzag" tubes, and for the general case (*m,n*), the term "chiral" tubes is used. The armchair CNTs are conductive for all the values of *m*, the zigzag CNTs behave as metals for *m=3q* (*q* is an integer) and as semiconductors for the other cases. Typically, CNTs are 0.1-10 $\mu$m in length, and their cross-sectional radius varies in the range 1-10 nm. More complicated structure is a multi-wall CNT, which may be considered as a set of coaxial single-wall CNTs of different types (*m,n*).

The mechanism of the CNT conductivity in THz and infrared frequency ranges is dictated by the intraband $\pi$–electron motion [196,200,231,232]. The conductivity for visible light is dominantly defined by the inter-band transitions of $\pi$–electrons between the valence and conductive zones [196]. The CNT conductivity is characterized by dramatically high anisotropy: the axial component of the conductivity strongly exceeds the transverse one [196]. It is important to note, that the intraband conductivity does not lead completely to the Drude mechanism. A broad and strong terahertz conductivity peak in the THz conductivity spectra has been universally observed in various types of SWCNT samples, containing both semiconducting and metallic nanotubes [199,233–245]. The physical origin of this peak was controversial for a long time, while it was important for ultrafast electronics and optoelectronics devices and, in particular, for different antenna applications. Considerable efforts were made to clarify its physical mechanism, both theoretical and experimental [199,245]. There were optical, THz, and DC transport measurements on highly enriched metallic and semiconducting CNT films. As a result, it was concluded that the broad and strong terahertz conductivity peak behaves consistently with the plasmon resonance explanation, firmly ruling out the other alternative explanations, such as the absorption due to curvature-induced gaps.

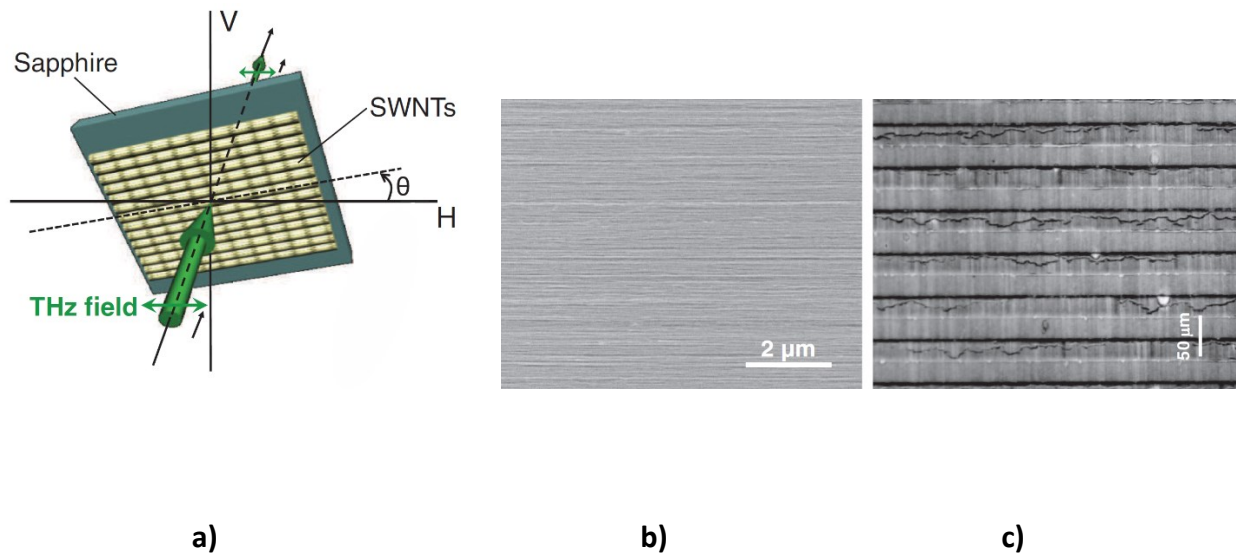

a)                          b)                         c)

**Figure 16.** a) Experimental configuration[231]. b) Scanning electron microscope image of a single-wall CNT (SWCNT) film sapphire substrate. c) Optical microscope image of a SWCNT film showing the high unifc of the CNT lengths. Copyright (2013) American Physical Society.

As a result of the recent investigations, it was established that different CNT structures are able to act as an antenna: the isolated CNTs[230,246–250], the CNT-bundles[230], the aligned CNT- arrays vertically oriented with respect to the growth substrate[231] (see **Figure 16**). The main shortcomings of the experimental samples at the previous steps lie in that they were grown by various methods and placed in a variety of THz transparent polymer films. Also, most of the samples consisted of randomly oriented bundles of both semiconducting and metallic nanotubes with a wide distribution of the lengths and diameters. The synthesis of the SWCNT samples with a high degree of alignment and high uniformity of the CNT length is performed by chemical vapor deposition[251–255]. It was applied to the CNT-antenna implementation[231].

Another approach of the antenna implementation is the preparation of the $SiO_2$-CNT structures by means of clinging of a CNT dispersed in water to previously silanized cladding-free silica fibers[256]. The CNT-coated fibers with the diameter around 130 $\mu$m and the thickness of the coatings of 20 $\mu$m were attached to a dielectric substrate. The composition consisted of 1/3 metallic tubes and 2/3 semiconducting ones. The contacts, keeping the distance between them at 1.5 mm, were made of silver paint. The refraction index and the conductivity of the chosen composite material contained two peaks in the THz frequency range due to the antenna resonances of surface waves in the SWCNTs[256].

The great potential of the CNT-structures as antennas en masse appears due to their strongly controllable value of photonic state density, which exhibits strongly impressed resonant behavior[197,198,257]. As a result, the Parcell effect strongly exhibits itself in CNTs. It opens one of the promising ways of the quantum antenna implementation based on the CNT excitation via the spontaneous emission of atoms placed in the vicinity of the CNT. The effective way of controlling the antenna properties of CNTs is the substitutional doping of bundles and metamaterials, containing a mixture of conducting and semi-conducting SWNTs, by either nitrogen or boron atoms[258]. As it was shown in Reference[258], the frequencies of the antenna resonances of the doped bundles are blue shifted with respect to the resonance frequencies of the undoped ones. The slightly attenuated plasmons and the antenna resonances due to the edge reflections exist for not-too-thick multi-walled CNTs in the far-infrared and the mid-infrared frequency ranges[259].

### 5.3 Quantum dots and quantum dot arrays

#### 5.3.1. Single QD dipole antennas

A promising class of nanoantennas is based on the semiconductor nanostructures[260]. The configuration of the experimentally implemented dipole antenna based on a semiconducting nanowire is shown in **Figure 17**[225,226,261]. In Reference[261], it was obtained the self-assembled InAs/GaAs QDs in etched photonic wires, or for flat QDs in zinc-blende III-V nanowires. It was shown that the designed structure may act as a quantum nanoantenna excited by the spontaneous emission of the QD with a highly directed radiation diagram. It is important to note that rather large efficiency is achieved over the remarkably broad spectral range, λ=70 nm at λ=950 nm.

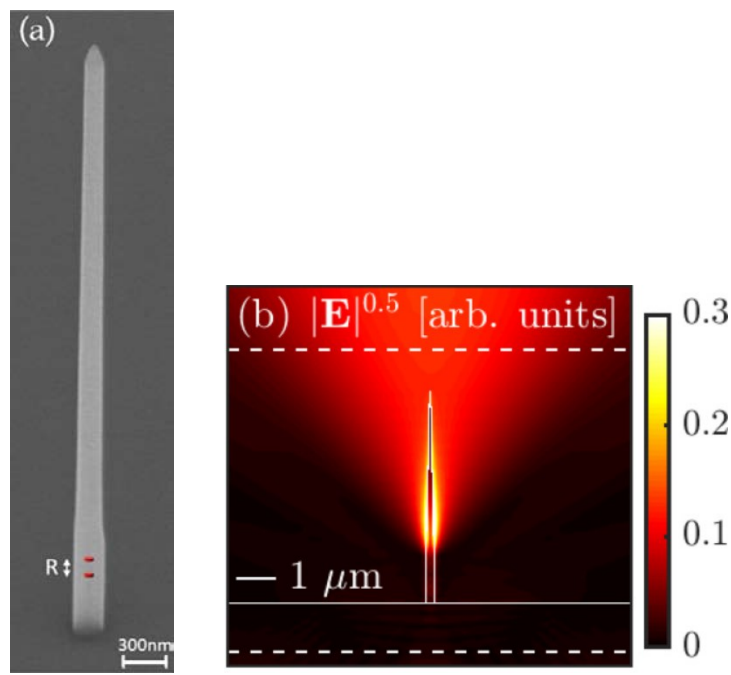

**Figure 17.** The tailored nanowire waveguide geometry [262]. (a) SEM image of a tapered InP nanowire waveguide. The position of the QD in the waveguide is schematically indicated by two red disks. b) The respective magnitude of the emitted electric field in the plane parallel to the tapered wire. Copyright (2019) American Physical Society.

In Reference [225], in order to enhance the antenna efficiency (light-extraction efficiency) of the QD emitter in the tailored nanowire, a bottom-up growth approach through independent control of both the nanowire shape and the QD location was used. Under the appropriate growth conditions, a single InAsP QD was placed exactly on the axis of an InP nanowire waveguide with a very small tapering angle towards the tip ($\alpha = 2°$). The used bottom-up approach allowed one to reach the near-unity antenna efficiency (97%, see Reference [225]), assuming a perfect mirror below the nanowire, a 1° nanowire taper and high coupling of the quantum emitter into the fundamental waveguide mode (compared with the higher-order modes with the coupling factor of 0.95, see Reference [225]) for the QD emitter on the axis of the nanowire waveguide.

Broadband optical antennas on QDs have been successfully used for the generation of entangled photons. An optical antenna with the extraction efficiency of 65% ± 4% was implemented, and a highly efficient entangled-photon source was demonstrated by collecting strongly entangled photons (fidelity of 0.9) at the pair efficiency of 0.372 ± 0.002 per pulse[263].

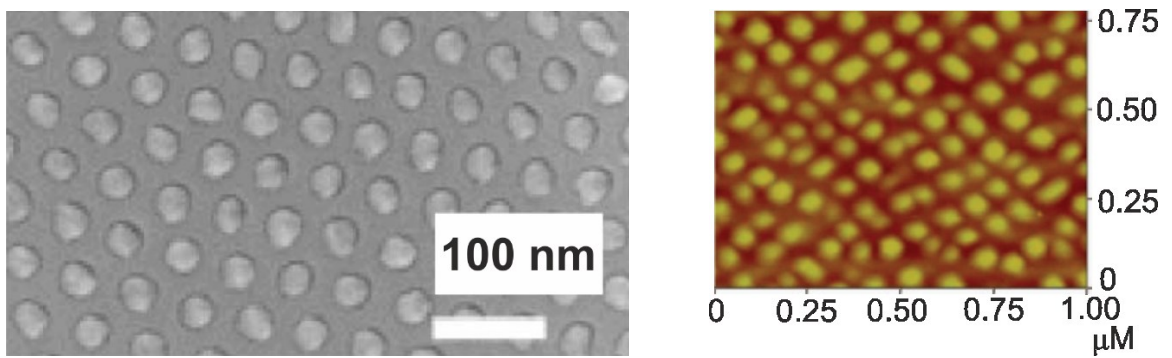

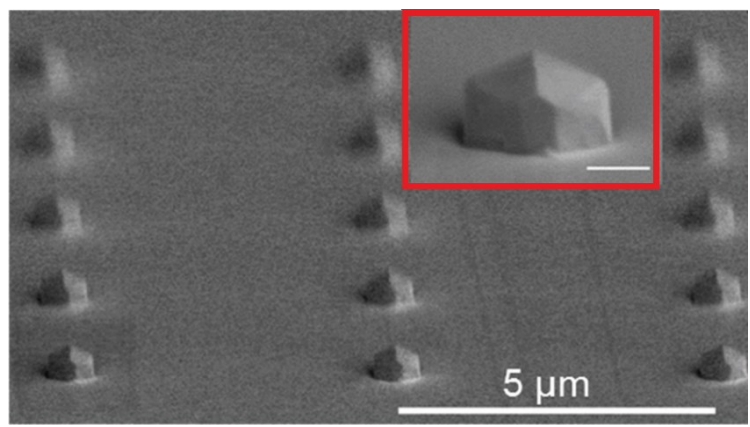

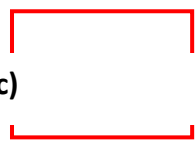

c)

**Figure 18.** The images of some types of the fabricated QD arrays: a) scanning electron microscope top-view image of the hexagonally ordered InAs quantum dot (QD) array [264] (Reprinted from Reference [264], with the permission of AIP Publishing); b) atomic force microscope image of the sample used in the work [137] (Copyright (2019) American Physical Society) ; c) 60° tilted scanning electron microscope image of the substrate-encoded size-reducing epitaxy-grown InGaAs mesa top QD array with the mesa top residing on the top of each nanomesa . The inset in the red frame is a magnified image of the mesa top single QD-bearing single nanomesa (scale bar of 300 nm) [265]. (Reprinted from Reference [265], with the permission of AIP Publishing).

### 5.3.2. QD arrays

To realize the 2D array of QDs feasible for the optical antenna applications, one should be able to fabricate a QD ensemble with a controlled packing density and a high degree of size, shape, and spacing uniformity. The analysis of the recent technological achievements shows that the novel procedures to overcome these limitations have been proposed and high-quality antenna arrays covering large areas have been fabricated [264–266].

For example, there are the reported highly ordered InAs QD arrays grown by molecular-beam epitaxy on nonlithographically nanopatterned GaAs [264]. There, the directed-growth method was used, which combines nonlithographic pattern transfer (scaled to large areas) with selective nucleation of Stranski–Kranstanov islands, thus combining high QD density, high crystal quality, size uniformity, and spatial periodicity. Approximately 20 billion dots are grown in 1cm$^2$ area with a small size dispersion, and forming a lateral metasurface in a hexagonal dense packing form (see **Figure 18a**). In particular, these techniques presage a pathway to the controlled growth of periodic QD metasurfaces, which offer macroscopic spatial coherence in the Rabi-wave propagation.

The sample produced in Reference [137] was a single layer of the self-assembled QDs fabricated by molecular beam epitaxy on a GaAs (311)$B$ substrate. After the 360-nm growth of $Al_{0.17}Ga_{0.83}As$ as a buffer layer, $In_{0.4}Al_{0.1}Ga_{0.5}As$ QDs were grown by the Stranski-Krastanov growth mode, and then a 200-nm $Al_{0.17}Ga_{0.83}As$ layer was grown as a capped one (**Figure 18b** shows an atomic force microscope image of the uncapped sample). One can see from **Figure 18b**, that the QDs in the ensemble have a lens shape with the diameter at the bottom of the QD $d$ = 50–70 nm and the height $h$ = 4–7 nm, with the QD density $1.1 \times 10^{10} cm^2$. The PL-peak emitted from the exciton ground state was located at 1.565 eV and exhibited the inhomogeneous broadening of 40 meV arising from the size distribution of the QDs.

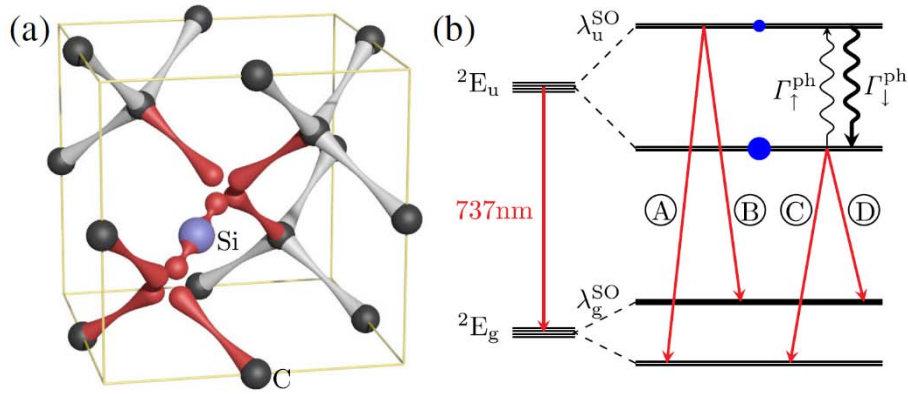

**Figure 19**. Electronic structure and optical transitions of the SiV center. (a) The center is aligned along the <111> axis of the diamond host crystal, with the silicon atom (Si) located in the middle of two empty lattice sites. (b) The scheme of the optical transition between the states $^2E_u$ and $^2E_g$. The spin orbit interaction ($\lambda_u^{SO} \sim$ 250 GHz, $\lambda_g^{SO} \sim$ 50 GHz) partially lifts the degeneracy giving rise to the doublets in the ground and excited states[204]. The transitions A, B, C, D are all dipole allowed. Copyright (2019) American Physical Society.

For the single photon sources array, it was reported about the triggered single-photon emission from GaAs(001)/InGaAs single QDs grown selectively on the top of nanomesas using the approach of substrate-encoded size-reducing epitaxy (see **Figure 18c**)[265]. As it was shown, it satisfies such an important requirement as the on-chip integrability with a multifunctional light unit, comprising an antenna, a waveguide, a cavity, a beam splitter, etc.

**5.4 Color centers**

Recently, the color centers have become quite popular quantum emitters systems for a broad range of applications ranging from quantum optics to quantum sensing. The color centers are crystallographic defects which occur, for example, when an anionic vacancy in a crystal lattice is occupied by one or more unpaired electrons[205]. The term "color center" was given due to the propensity of electrons in the vacancy to absorb light in the visible spectrum and make usually transparent material colored. The color center can exhibit exceptional brightness and room temperature operation. They allow for optical control of electrons and nuclear spins, and have sub-nanometer dimensions. The color centers can be formed naturally in the process of crystal growth, or be produced, for example, by a chemical vapor deposition process, by ion implantation or by irradiation of the crystal by neutrons or electrons, or even bright light pulses [204-210].

Two kinds of materials are commonly considered for hosting the color centers, which are perspective for photonics applications (and for constructing an antenna): diamond and silicon carbide[210]. The most extensively studied color center in diamond is the nitrogen-vacancy (NV) color center[209]. It was considered as one of the best possible candidates for room-temperature quantum registers build upon NV electronic, N and $^{13}$C nuclear spins due to very long spin decoherence times (which was experimentally demonstrated[267,268]), and the possibility of the optical read-out. The NV color center was also actively applied as a single-photon source and used as a convenient tool for addressing the fundamental issues in quantum and condensed matter theory, such as quantum teleportation and quantum process tomography[209]. However, the probability of coherent photon emission from the NV center is rather small, and its spectral stability is also problematic[269]. Recently, it has been found that the other color centers, in particular, the negatively charged silicon-vacancy (SiV) center (schematically shown in **Figure 19**), can be much better single-photon sources[204]. Supperradiance was experimentally demonstrated both with the NV and the SiV color centers[270,271].

Silicon carbide can also host the color centers potentially useful for photonics and quantum informatics[210,272]. Curiously, the silicon carbide was implemented in electronics for quite a long time. However, only recently, its color centers have been researched at the nanoscale[272]. It was shown that one can produce a scalable array of room-temperature single photon emitters or spin-qubits on the basis of the color centers in silicon carbide[273].

### 5.5. Antennas based on metal particles and metamaterials

### 5.5.1. Plasmonic antennas

This type of antennas is based on the surface plasmons and plasmonic resonances in nanoparticles of different configurations[43]. In fact, we touched on this topic in Section 5.1 with respect to the CNT-antennas. For the fabrication of metallic nanoparticles for quantum antennas, the so-called top-down and bottom-up techniques have been experimentally adapted[43]. The top-down approaches, e.g. electron beam lithography and focused ion beam milling, typically start from a thin multi-crystalline metal film on the top of an optically transparent but electrically conductive substrate. In general, the top-down approaches are able of fabricating large arrays of nearly identical nanostructures with well-defined orientations and distances[43]. On the other hand, the bottom-up approaches take advantage of chemical synthesis and self-assembly of metal nanoparticles[43].

The control over the interaction between the photons and the individual emitters is an outstanding problem in quantum antenna engineering. In particular, the principle of a cavity-free, broadband approach for engineering of the photon–emitter interactions via sub-wavelength confinement of optical fields near the dipole nanoantenna was demonstrated in Reference[48]. In this work, a single CdSe QD was optically excited in close proximity to a silver nanowire. The spontaneous emission from the QD coupled directly to the guided surface plasmons in the nanowire, causing the wire ends to light up. The results show that efficient coupling is accompanied by more than 2.5-fold enhancement of the QD spontaneous emission[68]. Notice that the effects of strong emitter-field coupling at room temperatures were experimentally demonstrated in plasmonic nanocavities[274].

The fabricated plasmonic antennas of the different configurations are presented in Reference[43]. For example, as it was shown in Reference[275], a small nanosphere excited by an optical field is able to behave as an effective ''nanocapacitor'' or a ''nanoinductor'' if the sphere is made of nonplasmonic or plasmonic materials, respectively. The imaginary part of the material permittivity may provide an equivalent nanoresistor. These concepts indeed provide new possibilities for miniaturization of quantum electrical circuits and antennas operating at the optical frequencies. It is important to note, that for sharp corners and nanogaps, the near field can be strongly (orders of magnitude) enhanced due to the increase of local photonic state density. The same effect is possible in a plasmonic optical dimer nanoantenna along with high levels of the radiation efficiency and rather weak absorption in the optical range[275]. The systematic analysis of the contribution of various plasmonic effects in the metal–semiconductor (molecular)–dielectric arrangements for using in different types of practical devices is given in Reference[276].

Among the properties of the nanomaterials for nanoelectronic applications, the chemical stability is also a relevant issue. Ag and Cu are known to quickly corrode under the ambient conditions (formation of oxides and sulfides), while Al is able to form thin passivation films of $Al_2O_3$. Au is the plasmonic metal that is used experimentally, since it combines favorable optical properties in the red and near-infrared frequency ranges with high chemical stability[277]. Carbon-based nanomaterials are better conductors than metals at the diameters smaller than 5 nm and are characterized by the excellent chemical stability[278]. Therefore, the carbon materials, such as graphene and nanotubes, may become the building blocks for optical antennas. Certain prospects have been opened by the search for new materials with the plasmonic properties. For example, the study of plasmonic excitations in one-atom-thick sodium chains is given in Reference[279].

### 5.5.2. Metamaterial nanoantennas

One of the steps forward to an efficient design of quantum antennas is using of the novel types of metamaterials, such as "epsilon-near-zero" (ENZ), "epsilon-mu-near-zero" (EMNZ) media ("zero-index" media) and the left-handed materials (LHM). Such types of media are proposed as an alternative platform to manipulate the decay of quantum emitters, possibly leading to efficient usage in the quantum antennas control[45,280–282]. A thin layer of ENZ material demonstrates the existence of Ferrell-Berreman modes with a shallow slope within the frequency bands bounded by the points of zero group velocity[283,284]. Since these "stop-light" modes are formed in a film thinner than the skin depth, the transverse field is expected to be homogeneous throughout the film. The multiple arrays of gold antennas on the indium tin oxide (ITO) substrate were fabricated and experimentally studied in Reference[284]. The strong modification of the linear response of plasmonic antennas (in particular, the independence of the resonant wavelength on the linear sizes of the dipole antenna) was demonstrated.

The scheme, which is able to produce a highly directional single photon with almost 100% efficiency was proposed in Reference[285]. It is based on the quantum emitter placed inside a Fabry-Pérot cavity for which every wall consists of both LHM and ZIM layers. As it was noted in Reference[285], such type of an antenna is promising for the applications in quantum communication and optical quantum computing.

## 5.6. Quantum antennas and quantum networks: loading and matching

The main functions of any antenna include connecting a feeding source, or a receiving detector or a transmission line in order to transmit / receive a given signal to / from the far field. In this sense, the problem of the antenna matching with a source/detector is of fundamental importance for a proper design and use of quantum antennas [5,286]. One of the widely used approaches for the matching analysis in classical antennas is based on the concept of Thevenin or Norton equivalent networks [5]. These approaches are conceptually equivalent and lead to identical results. Therefore, we will limit our consideration by the network of a Thevenin type (see **Figure 20**) for a receiving antenna (a transmitting antenna may be considered via the reciprocity theorem [5,35]). This model manipulates with the complex impedances of the antenna and load $Z_A = R_A - iX_A$, $Z_T = R_T - iX_T$, with the antenna resistance given by $R_A = R_r + R_L$, where $R_r, R_L$ are the radiation and the loss resistance of the antenna, respectively [5]. Following this picture, the powers delivered to the impedances $R_T, R_r, R_L$ are $P_T = |V_T|^2 R_T / 2|Z|^2$, $P_r = |V_T|^2 R_r / 2|Z|^2$, $P_L = |V_T|^2 R_L / 2|Z|^2$ where $V_T$ is the peak voltage induced by the incident field, $Z = R_r + R_T + R_L - i(X_A + X_T)$ is the total impedance. The equivalent Thevenin network may be generalized for quantum antennas via the exchange $|V_T|^2 \to \pi^{-1} \int_{-\infty}^{\infty} \langle \hat{V}^{(-)}(0) \hat{V}^{(+)}(\tau) \rangle e^{i\omega\tau} d\tau$, where the symbols $(\pm)$ mean the positive/negative frequency components of the voltage operator. Therefore, the maximal power at the load happens, as in the classical case, only when we have conjugate matching between the antenna and the load, which is

$$R_r + R_L = R_T; \quad X_r = -X_T.$$
(62)

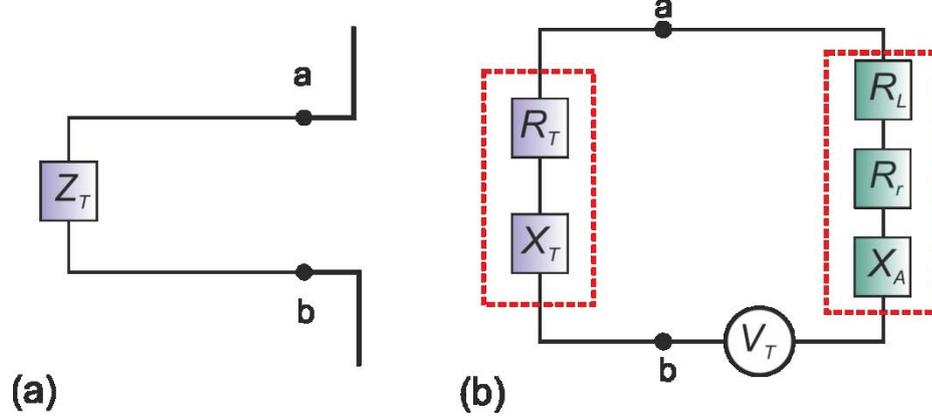

**Figure 20**. a) Dipole antenna, loaded with the impedance $Z_T = R_T - iX_T$. b) Its Thevenin equivalent network. Antenna terminals are denoted as *a-b*.

However, the physical models of such values as the input impedance at the feeding point, the antenna radiation resistance, and the load impedance are distinct from conventional classical analogues in the radio-frequency range. The dispersion of metals at infrared and optical frequencies is rather high and cannot be ignored. Inspired by the conventional radio-frequency antenna concepts[5], the concepts of the optical input impedance, the optical radiation resistance, loading, and the impedance matching have been developed in the optical domain for classical light[287,288]. In References[287,288], these concepts have been applied to plasmonic nanoantennas in the form of nanodipoles. Particularly, in the range of negative permittivity, the network elements exhibit the inductive (capacitive) frequency dependence of the reactance, with the opposite sign of the reactance with respect to the ordinary inductors (capacitors). This is the reason to speak about "effective" negative values of inductances (capacitances), which have no analogues in ordinary electric circuits[287].

As an efficient theoretical framework for the quantum antenna analysis, the theory of electric circuits with quantum emitters was proposed[289]. In order to bridge the classical electric circuit theory with the quantum theory of nanoobjects, the quantum emitters/detectors inside nanodevices were described by their effective admittances. The effective admittance, essential for quantum nature, is defined as the general susceptibility and calculated using the Kubo-technique[289]. In order to account for the decoherence, the concept of an effective Hamiltonian was used, with the decoherence rates taken from the experimental data[290,291]. To illustrate the general concept, the electromagnetic crosstalk between two identical circuits with identical two-level emitters in entangled states coupled via d-d interactions was analyzed. The circuit was described in terms of an equivalent symmetric two-port, and it was shown that the combination of the entanglement and the d-d interactions enables a dramatic change in the scenario of the crosstalk. For example, in contrast with the classical crosstalk (see Reference[292]), the excitation induced in one of the ports will be redistributed in equal parts between both ports, in spite of rather small inter-emitter coupling[289].

An optical nanodevice consisting of a receiving as well as an emitting nanoantenna, connected by a transmission optical line was studied in Reference[286]. The possibility of matching the nanoantenna impedance to the transmission line was shown, and it was clearly demonstrated that the systems of interconnected nanooptical elements can be optimized by applying the concept of the impedance matching, conventional in classical antennas.

One of the steps forward to an efficient design of quantum antennas is the usage of the novel types of metamaterials, such as "epsilon-near-zero" and "epsilon-mu-near-zero" media ("zero-index" media) for conjugative matching. Such types of media are proposed as an alternative platform to manipulate the decay of quantum emitters, possibly leading to an efficient usage in quantum antennas control[280,281].

## 5.7. Electromagnetic compatibility (EMC) at the nanoscale

Since the early days of radio and telegraph communications, it has been known that a numerous sources of electromagnetic emission cause unwanted crosstalk or noise in various electronic and electrical devices such as radio receivers and telephone communications[293]. Radar transmitters also transmit pulses of a single-frequency carrier. As this carrier frequency is pulsed on and off, these pulses are radiated out of the antenna, strike a target, and come back to the radar antenna. The total transit time of the wave is related to the distance between the target and the radar antenna. The spectral content of the radar pulses is distributed over a large band of frequencies around the carrier.

An electronic system, that is able to function compatibly with other electronic systems and not produce the interference, is defined to be electromagnetically compatible with its environment. Antennas are obviously the major ingredient in the discipline of EMC. The intentional antennas, such as the radar antennas, generate electromagnetic fields that couple to electronic devices and create the susceptibility problems [293]. A system is defined as electromagnetically compatible with its environment, if it satisfies three criteria: i) it does not cause electromagnetic crosstalk with other systems; ii) it is not susceptible to electromagnetic emissions from other systems; iii) it does not cause interference with itself.

The present-day developments in nanoscience made quantum entanglement relevant for the proper operation of nanoelectronic and nanophotonic devices. Recently, the quantum entanglement has been identified to be a promising tool facilitating the growth in the level of integration and the reduction of the operation power in nanoelectronics and nanooptics. Among the devices based on the entanglement are the qubits of various types (see References[294–296]) and the quantum amplification by superradiant emission of radiation (QASERs) (see Reference [297]). The electromagnetic interference at the nanoscale due to the d-d interactions may be accompanied by the formation of the undesirable quantum entanglement "antenna-antenna" or "antenna-circuit". As a mechanism of the quantum interference, the entanglement can produce a combined "electromagnetic-quantum" crosstalk and induce long-distance and long-living electrical correlations [289]. As a consequence, the qualitative picture of the crosstalk at the nanoscale is dramatically changed, and the EMC-concept conventional for macroscopic radio-frequency devices needs revising. As it was shown in References [289,298], the electric crosstalk becomes anomalously high even for rather weak mutual penetration of electromagnetic field. The manifestation of the entanglement can be strongly influenced by various elements of the environment (for example, resonant cavities and waveguides near a cutoff). Such long-distance entanglement was experimentally observed on a set of trapped ions at inter-ion separations that are much larger than the wavelength of the resonant transition [299]. The role of quantum correlations in the nano-EMC can be dual. Depending on the system parameters, they are able to open additional unwanted channels of anomalous crosstalk. On the other hand, they can provide the tools for electromagnetic crosstalk suppression. Examples of both types have been analyzed in detail. [289]

## 6. Outlook

In this review, we have followed bringing together the classical antenna concept, historically stemming from the radio-frequency applications, and the quantum concept of the emitter/detector interaction with the field, considered as a reservoir of bosonic modes. Notice that even when the emitter dynamics is an essentially quantum process, it does not mean that one mandatory needs quantum description of the emitted light. The role of the reservoir structure may be and quite often reduced just to modification of the spontaneous decay rate of the emitters' excitation and to the frequency shift. Also (and that is in a classical antenna spirit, too), the field distribution both in the near-field and in the far-field zones can be designed by appropriate structuring of the emitters environment, and by designing the emitters properties. Notice that the field state may be essentially quantum (like the single-photon state, which certainly is).

However, as explained throughout this review, when one has a set of interacting quantum emitters, or an entangled initial state of the antenna, or non-classical driving, or the reservoir structured to such an extent as to distinguish one, or several modes strongly interacting with the emitter, one unavoidably enters the kingdom of quantum and needs the full-pledged quantum description of the emission/absorption and the field propagation process. It is still possible to generalize the classical description and introduce the concepts like the characteristic impedance and the radiation pattern (such patterns can be introduced, actually, not only for field and power, but for correlation functions of different orders). But for quantum systems, the borders between the feeds, the emitters and the radiation reservoirs may be rather vague. Interactions may correlate the emitters, the feed sources and the parts of the reservoir and turn them into a single compound quantum emitter (like, for example, a chain of coupled emitters in a waveguide).

Nowadays, nanoscale antennas are very intensively researched and applied for shaping light at THz, infrared and visible ranges. The emerging field of quantum antennas aims to provide the tools for shaping not only field amplitude/intensity distribution, but to shape also field correlations and quantum features of the field states in the far-field zone. Thus, one can envisage the implementation of the quantum antennas for far-field quantum enhanced sensing, for measurements and metrology, for super-resolution imaging, for quantum communication and information processing. There already exist the schemes allowing one to reach the super-resolution in determination of the target parameters, to shape the ultra-narrow radiation beams, to demonstrate the feasibility of quantum radars. It can be expected, that the quantum noise and dissipation (instead of being the unwanted factors suppressed by all means in classical devices) will transform at the quantum level to working resources of information processing, paradoxical as it sounds. Of course, to accomplish this task, it will be required to further improve the antennas configuration, as well as their matching with feeding sources (or detectors) when combined together. It will be reachable on the basis of new types of materials and metamaterials, as well as on the future progress in nanotechnologies. Of course, the quantumness brings about new problems. Entangled states are considerably more fragile than classically correlated states, quantum antennas are more difficult to create, protect from losses and feed. One cannot easily bring vast knowledge on the design and functioning of classical antennas into the quantum domain. Designing the antenna geometry, interaction strengths, initial quantum states and quantum field characteristics are formidable nonlinear problems. Nevertheless, the field of quantum antennas is rapidly developing. One must hope that these still exotic devices will soon find their proper places both in the experimental lab and in the line of practical commercial implementations.

**Acknowledgements**

The authors gratefully acknowledge the support from Tel Aviv University. D. M. also acknowledges support from the EU Flagship on Quantum Technologies, project PhoG (820365). G. S acknowledges support from the H2020, project TERASSE 823878.